\documentclass[aps,prb,twocolumn,groupedaddress,showpacs,longbibliography]{revtex4-1}

\usepackage{graphicx}
\usepackage{amsmath,amssymb,textcomp,epsfig,indentfirst,color}

\begin{document}
\title{Segregation of impurities in GaAs and InAs nanowires}
\author{M. Galicka}\affiliation{Institute of Physics, Polish Academy of Sciences, Al. Lotnik\'ow 32/46, 02-668 Warsaw, Poland}
\author{R. Buczko} \affiliation{Institute of Physics, Polish Academy of Sciences, Al. Lotnik\'ow 32/46, 02-668 Warsaw, Poland}
\author{P. Kacman} \affiliation{Institute of Physics, Polish Academy of Sciences, Al. Lotnik\'ow 32/46, 02-668 Warsaw, Poland}

\email{galicka@ifpan.edu.pl}

\begin{abstract}
Using {\it ab initio} methods based on the density functional theory, we investigate the segregation and formation energies for various dopants (Si, Be, Zn, Sn), commonly used to obtain p- or n-type conductivity in GaAs and InAs nanowires. The distribution of Au and O atoms, which may be
unintentionally incorporated during the wire growth, is also studied. The calculations performed for nanowires of zinc blende and wurtzite structure show that the distribution of most of the impurities depends on the crystal structure of the wires. For example, it is shown that the same growth conditions can lead to lower energy for Si substituting Ga (donor) in the wire of wurtzite structure and substituting As (acceptor) in the wire of zinc blende structure. In contrast, we obtain that gold and oxygen atoms always tend to stay at the lateral surfaces of GaAs and InAs nanowires, in agreement with experimental findings, while for beryllium the lowest energies are found when the impurities are located in sites in the center of the wurtzite wire or along the $[1,0,\overline{1}]$ axis from surface to the center of the zinc blende wire, what can explain the recently observed diffusion of this impurity into the volume of GaAs wires.
\end{abstract}

\maketitle

\section{Introduction}
Successful realization of most of electronic nanodevices requires a high carrier concentration so
as to reach sufficient conductivity. The excellent properties of III-V semiconductors, in particular
the high electron mobility which can be obtained in GaAs and InAs, turns nanowires (NWs) from these
materials into natural candidates for building blocks of such high speed applications. The properties
of semiconductor NWs differ from those of their parent bulk materials due to quantum
confinement and surface related effects -- pure semiconductor NWs usually have low concentration of
carriers. Thus, the application of semiconductor NWs in novel electronic devices requires controllable
p-type and n-type doping. This task still remains a big challenge to the growers, despite substantial
efforts, both experimental\,\cite{Piccin,Tambe,ThelanderNanotech,Ford} and theoretical.
\cite{Leao,Rurali,peelaers-apl,Ghaderi,dosSantos,Shu11,Shu12} Most of the publications describing doping
of III-V NWs focus on InAs NWs.\cite{ThelanderNanotech} In these structures the surface Fermi level is
pinned in the conduction band. This makes n-type conductivity in InAs NWs easy to obtain at the expense
of difficulties in p-type doping.\cite{Hasegawa} In contrast, in GaAs NWs the Fermi level at the surface
is pinned approximately in the center of the band gap and both, controlled p- and n-type doping, can be
produces. Doping of GaAs NWs grown by molecular beam epitaxy (MBE) has been demonstrated in different
means. Czaban {\it{et al.}} used Be and Te as p- and n-type dopant precursors respectively,\cite{Czaban}
while Fontcuberta i Morral {\it{et al.}} pointed out that Si may act as both, acceptor and donor, by just
changing the operating temperature during growth as shown before for 2D epitaxial growth of GaAs.
\cite{Colombo,Dufouleur} The incorporation of Si and Be into GaAs NWs was investigated in a further study.
\cite{hilse-si} Recently, the use of tetraethyl tin as a precursor material for successful n-type doping
of GaAs NWs, grown by metal organic vapor phase epitaxy, was reported.\cite{Gutsche} Moreover, it occurred
that not only specific growth conditions but also the crystal structure of the wires may lead to different
incorporation behavior than known for planar layers.\cite{Dufouleur,Gutsche,Erwin,hilse-si} This was shown
in particular for the group IV element, silicon, which is known to be amphoteric in III-V compounds. Namely,
it was found that although under typical MBE growth conditions Si acted as n-type dopant for GaAs, p-type
conductivity was obtained in Si-doped GaAs NWs which adapted wurtzite (wz) structure in similar growth
conditions.\cite{Piccin}

As mentioned above, doping of NWs was studied theoretically by several groups. They focused, however, mainly
on silicon NWs (see Ref.\,\onlinecite{Chen} and the references therein). There are very few publications which 
describe theoretically, incorporation of non-magnetic
dopants in GaAs and InAs NWs. To our best knowledge, doping of GaAs NWs was considered only in Ref.
\onlinecite{Ghaderi}, where only Si doping of wz structure GaAs NWs was studied. In contrast, Cl\`{a}udia dos
Santos and co-authors considered Cd and Zn doping of InAs NWs having the zinc-blende (zb) structure.\cite{dosSantos}
In Ref. \onlinecite{dosSantos} the studied NWs were extremely thin (with diameters of 1~nm) and their lateral
surfaces were passivated with hydrogen atoms. Recently, the role of the surface dangling bonds and the effect
of molecular passivation on doping of InAs NWs was also studied by Haibo Shu {\it{et al.}}\cite{Shu11,Shu12}

In this paper we report a systematic theoretical study of several dopants, which are usually used to obtain
p- or n-type conductivity in GaAs and InAs NWs, i.e., Si, Be, Sn and Zn. In contrast to the previous papers,
we assume that the surface atoms are not saturated by foreign atoms. This assumption is justified by the fact,
that we consider incorporation of dopants during the growth process in high vacuum conditions, e.g., MBE.
To check what is the preferential location of a given dopant in the wire, i.e., to answer the question whether
it is possible to explain the problems of doping of III-V NWs by segregation phenomena, we study the energy
of the wire with the impurity situated in different sites. Such analysis proved already to be very useful,
e.g., for predicting conditions to obtain ferromagnetism in GaAs NWs with magnetic Mn ions.\cite{galicka2}
In Ref. \onlinecite{galicka2} it was shown that the distribution of Mn ions and thus the electric and magnetic
properties of (Ga,Mn)As NWs depend strongly on the crystal structure. Since the III-V semiconductor NWs can
grow in both, zb and wz structures, we consider here both types of wires to check whether the crystal structure
has an impact on the distribution of the studied impurities in GaAs and InAs NWs.
Finally, we consider the distribution of Au and O atoms in the studied NWs. Oxygen can be used for n-type doping
of GaAs\,\cite{Huang,Salem} but also unintentional incorporation of O atoms in GaAs wires has been reported in the
literature.\cite{Zhi} On the other hand, gold can defuse into the wire from the catalyst particle during
Au-assisted vapor - liquid - solid (VLS) growth. This is known to be a serious problem in Si NWs.\cite{Allen}
Although it seems that the unintentional incorporation of Au in GaAs and InAs wires is much lower\,\cite{Perea}
the measured Au doping values can make ballistic transport through
the nanowires difficult to obtain.\cite{Hadasgold} This makes the study of the distribution of Au atoms in the
wires very important. It should be mentioned here that to avoid Au incorporation self-catalyzed growth of GaAs
and InAs NWs has been pursued by several groups (see, e.g., \onlinecite{Mandl2006,Jabeen,AFMorral,Peter}).
\section{Methodology}
\begin{figure}[hbt]
 \centering
a)  \includegraphics*[width=0.45\textwidth]{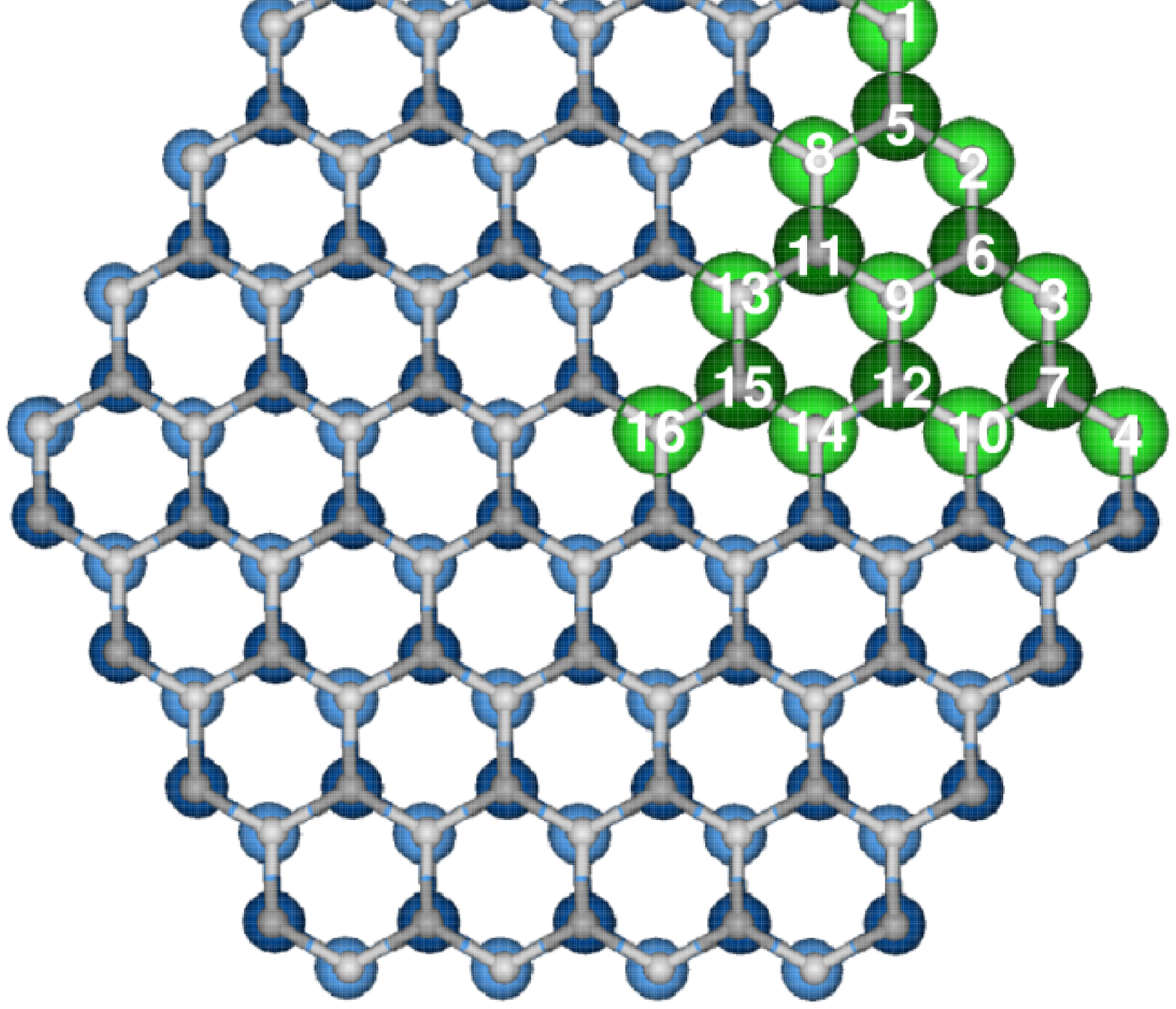}
b)  \includegraphics*[width=0.45\textwidth]{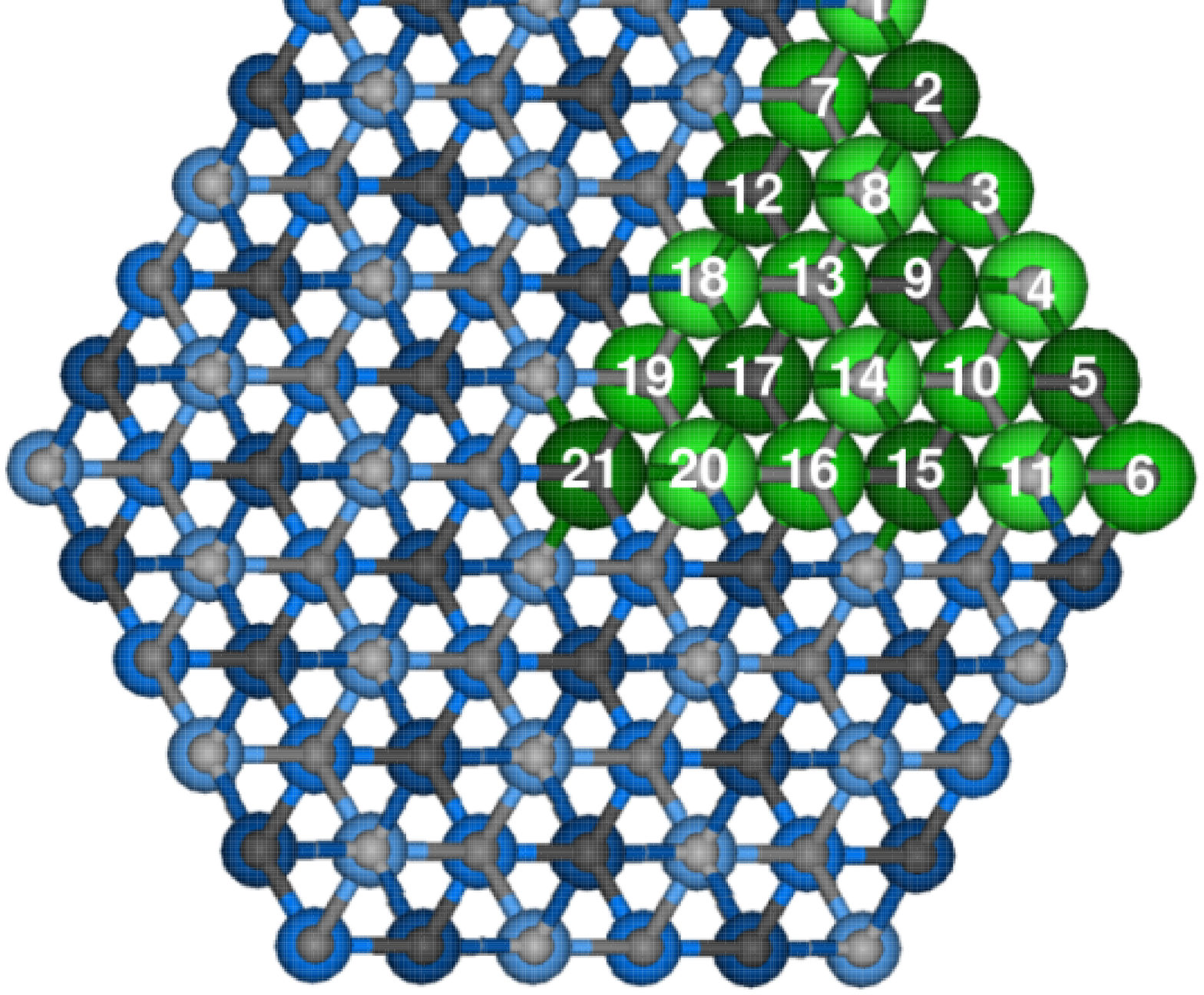}
\caption{(Color online) Cross section of an a) wz NW with diameter d=2.8~nm and b) zb NW (d=2.4~nm). The green balls show the 10 nonequivalent sites, which we consider as substituted by a dopant. The blue balls denote Ga(In) atoms, gray - As atoms. The darker colors denote atoms in the deeper double layer.}
  \label{fig:positions}
\end{figure}

In our calculations the doping of the wires is performed by substituting one cation or anion in the elementary
cell of the NW with an impurity atom. One foreign atom in the elementary cell of the studied NWs corresponds
to a concentration of impurities in the range from $\sim\,$0.5\% to $\sim\,$1.6\%, depending on the crystal
structure and diameter of the wire. We study the distribution of impurities by comparing the total energies of
the NWs with an impurity at various non-equivalent substitutional sites. In Fig.~\ref{fig:positions} we present
the dopant positions considered here. To a given cation site and the anion above this cation we ascribe the same
number. It should be emphasized that in the cross section corners, i.e., in the positions 1 and 6, the geometry
of bonds is different for cations and anions. Namely, in site 1 the cation has two dangling bonds while the anion
has one dangling bond, and vice versa in site 6.

The energies of the NWs are calculated by \emph{ab~initio} methods, based on the density functional theory (DFT).
The energy of the wire with the dopant occupying the site exactly at the center of the wire cross section is taken
as a reference point, i.e., as zero on the energy scale. The segregation energy is defined by the difference
between the energy of the NW with an impurity at a given site and the reference energy:
\begin{equation}{\label{eqn:segregation}}
E_S = E_{doped\ NW} - E_{center\ doped\ NW}.
\end{equation}
Thus, the negative value of the segregation energy for a dopant atom at a given site means that this impurity
prefers to substitute the host atom at this particular site rather than be at the center of the wire. In contrast,
positive segregation energy values denote the situation when substitution of an atom in the center of the NW will
be energetically more favorable. Let us mention that in a study of doped Ge wires,\cite{peelaers-apl} performed 
with use of a similar method, the values of formation energy, i.e., energy needed to insert the impurity atom 
(taken from a reservoir) into a given NW's site, from which the original atom has been removed (to a reservoir), 
were shown to depend strongly on the size of the simulated supercell and that segregation energy, being a relative 
quantity, is free of these systematic inaccuracies.
However, by the study of segregation energies we can only answer the question where an impurity prefers to stay within 
the wire. To answer the question for which impurity the energy cost of substituting a host atom is lower, one has to 
compare their formation energies, which depend on atomic and electronic chemical potentials.
The formation energy $\Omega$ of a neutral impurity $X$ can be calculated according to the formalism proposed by 
Northrup and Zhang, \cite{Northrup}
\begin{equation}{\label{eqn:formation}}
\Omega = E_{tot}[X] - E_{tot}[NW] + \mu_{Ga/As} - \mu_X,
\end{equation}
where $E_{tot}[X]$ is the total energy derived from the calculation of the nanowire with an impurity $X$, $E_{tot}[NW]$ 
is the total energy of undoped nanowire. The term $\mu_{Ga/As}$ denotes the chemical potential of the corresponding 
substituted Ga or As atom, while $\mu_X$ denotes the chemical potential of the impurity atom. Our analysis is performed 
with an assumption of thermal equilibrium conditions of the impurity region with the surrounding nanowire, i.e.,
\begin{equation}
\mu^{bulk}_{GaAs} = \mu_{Ga} + \mu_{As}.
\end{equation}
Using $\mu_i\leq\mu_i^{bulk}$ with $i\in{Ga,As,X}$ and the definition of the heat of formation of GaAs
\begin{equation}
\Delta H = \mu^{bulk}_{Ga}+ \mu^{bulk}_{As}- \mu^{bulk}_{GaAs},
\end{equation}
we can limit the range of possible values of $\Delta\mu$ to $-\Delta H\leq\Delta\mu\leq +\Delta H$. The quantity $\Delta\mu$ 
is the difference between the chemical potentials of Ga and As and is given by
\begin{equation}
\Delta\mu = (\mu_{Ga} - \mu_{As}) - (\mu^{bulk}_{Ga} - \mu^{bulk}_{As}).
\end{equation}
Thereby, the situation $\Delta\mu=+\Delta H$ corresponds to Ga-rich growth conditions, while $\Delta\mu=-\Delta H$ 
corresponds to As-rich conditions.

As mentioned above, we incorporate the dopants into GaAs and InAs wires of both wz and zb crystal structures.
The only considered NWs are those grown along [0001] direction (wz) and along [111] axis (zb), because a vast 
majority of III-V NWs grow only along these directions. It has been also shown theoretically that these growth 
directions and the types of facets presented in Fig.~\ref{fig:positions} constitute the most stable configurations 
of GaAs and InAs NWs.\cite{JPCM,Glas,dubrowski,Ciraci,Shtrikman}
It is well known that with \emph{ab~initio} methods only very thin NWs, much thinner than the real ones, can be simulated. This can lead to overestimation of the side surfaces' effects in the results, deceptive when such calculations are used to determine, e.g., the crystal structure of thicker wires. This problem is, however, not so important in our study, in which we try to answer only the question: is there any energy difference between the wires with an impurity situated at the side facets and with the impurity inside the core of the wire. The diameter test, in which we considered zb wires of several different diameters, 0.9~nm, 1.4~nm, 1.8~nm and 2.4~nm, has shown that for the two highest diameters the obtained segregation energies are practically the same. Similarly, the results do not depend considerably on the diameter in wz NWs with diameters from 2.0~nm to 2.8~nm. Congruous studies reported in Ref.~\onlinecite{Leao} also indicate that for diameters larger than approximately 3~nm the P impurity levels inside the Si NWs recover bulk characteristics. In the following, all the presented results are for the thickest wz NWs with diameters of 2.8~nm and the thickest (2.4~nm) zb wires.
As already mentioned, we assume that the surface atoms are not saturated by any foreign, for example, hydrogen atoms, because we study the process of NWs growth and incorporation of the impurities. Still, for the calculations of the density of states of already doped wires and discussion about the positions of the Fermi level, the side facets of the wires have been passivated by partial atomic charges of hydrogen atoms.

The calculations are performed using the Vienna \emph{ab~initio} simulation package (VASP).\cite{nw4,nw5} Each
of the initial structures is cut out from an appropriate bulk material, in either zb or wz structure.
The wz elementary cell is doubled along the $c$ direction, to guarantee that there is enough separation between
the impurity ion and its periodic images. Thus, the smallest supercell used in our study contains 122 atoms (zb)
whereas the largest (wz) 384 atoms. For different dopants a plane wave basis set is adopted with different cutoff
energies: for Si the cutoff energy was equal to 307~eV, for Be -- 375~eV, Sn -- 261~eV, Zn -- 346~eV, Au --
287~eV and O -- 500~eV. In each case all the energies and chemical potentials are calculated with these cutoff values.
These cutoff energies are sufficient to ensure that the results do not depend on the
size of the plane wave basis. The k-points are generated with a ($1\times 1\times n$) mesh, where $n\geq 50~$\AA$/c$
and $c$ is the unit cell dimension in the growth direction. The interaction between the valence electrons and
the ionic cores is included using the projected augmented wave method (PAW).\cite{PAW} The exchange correlation
energy is calculated within the generalized gradient approximation (GGA).
The atomic positions leading to minimum energy are determined with relaxation of all atomic positions as well
as of the unit cell dimension in the growth direction, and with  full reconstruction of the NW's surfaces.
In each simulation process the atomic coordinates are relaxed with a conjugate gradient technique. The criterion
that the maximum force is smaller than 0.01~eV/{\AA} is used to determine the equilibrium configurations. Neither
pressure nor temperature are included in our calculations.

\section{Results}
\begin{figure}[hbt]
 \centering
a)\includegraphics*[angle=-90,width=0.48\textwidth]{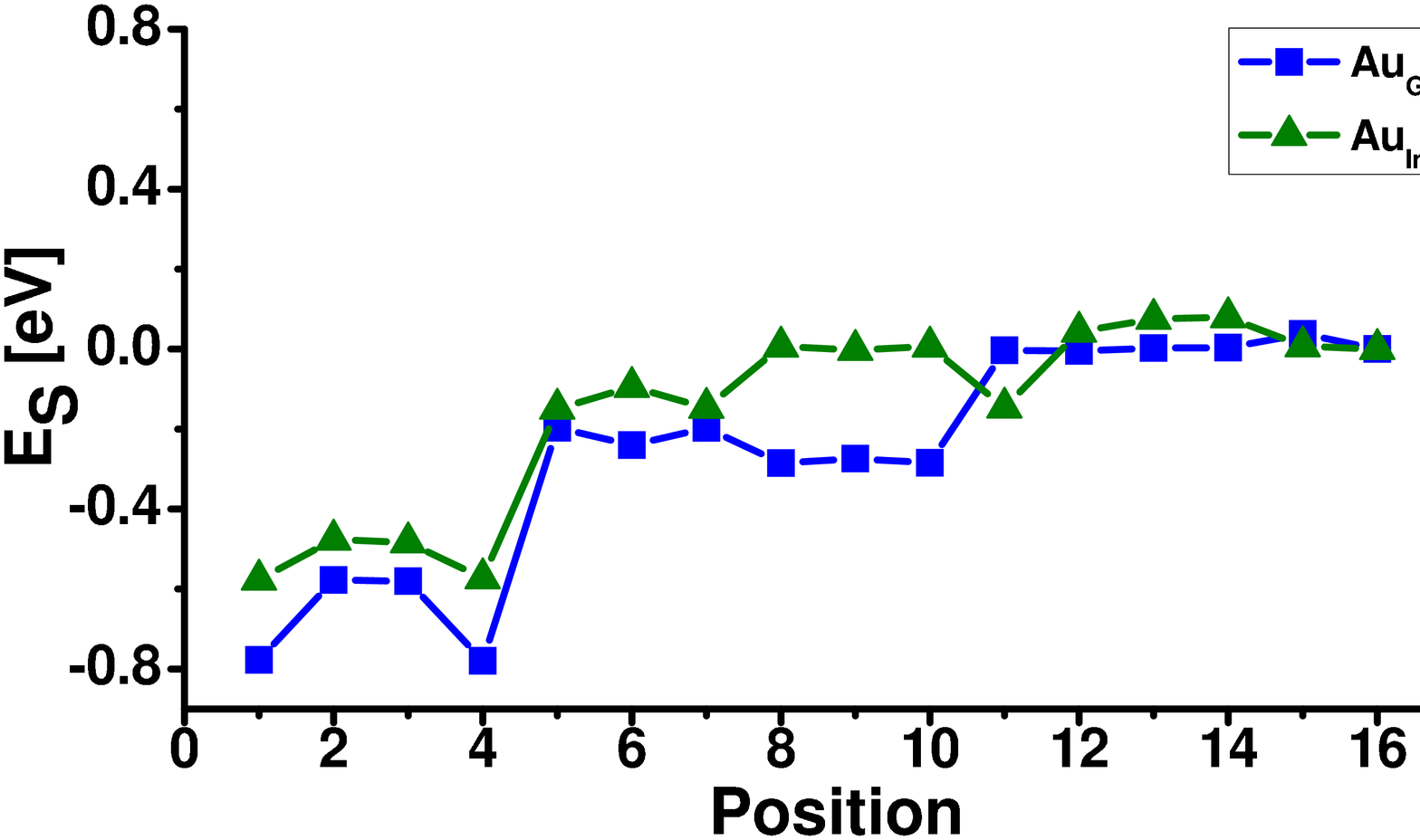}
b)\includegraphics*[angle=-90,width=0.48\textwidth]{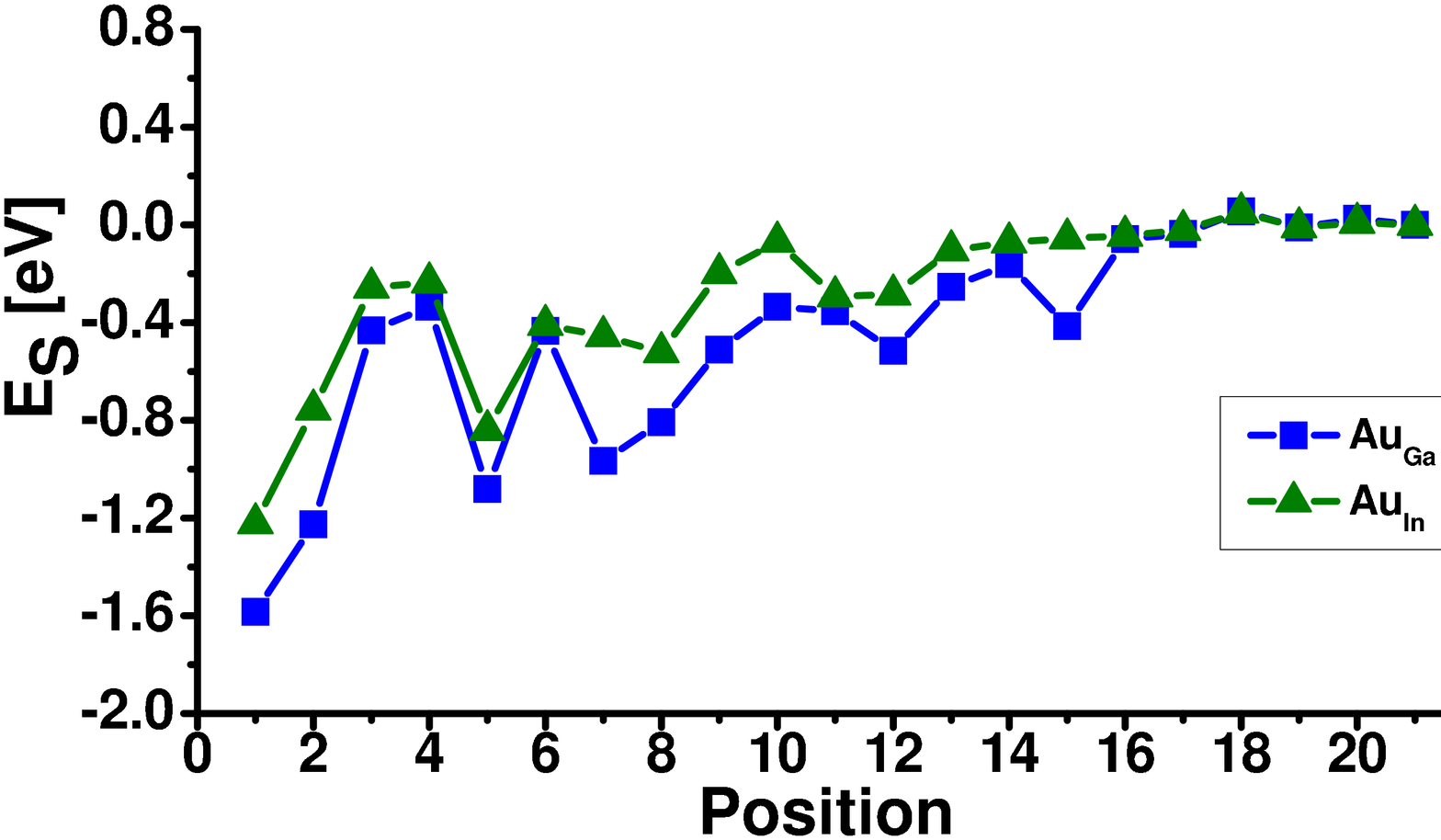}
\caption{(Color online) The segregation energy in a) wz (diameter 2.8~nm); b) zb (diameter 2.4~nm) InAs (green) and GaAs (blue) NWs as a function of different Au atom positions.}
  \label{fig:segregation_au}
\end{figure}

The distribution of gold in the cross-section of InAs wires was studied experimentally using a technique called
pulsed-laser atom probe tomography.\cite{Perea} In Ref.\,\onlinecite{Perea} it was shown that Au atoms accumulate
in a shell at the lateral surfaces of the wire. We start our study by considering Au impurities in InAs NWs to check
if our calculations can describe this experimental finding. We analyze InAs NWs with an Au atom in various cation
sites and focus on the search of the impurity positions for which the segregation energy is the most negative, i.e.,
the sites in which the energy cost of substituting the cation by an Au atom is lower than in the other nonequivalent
sites. In Fig.~\ref{fig:segregation_au} the results for both wz and zb InAs as well as GaAs NWs are presented. In
this and all following Figures the diameters of the wz NWs equal to 2.8~nm and of zb NWs to 2.4~nm. In all Figures
the calculated values of segregation energy are connected by a line to guide the eye.

\begin{figure}[hbt]
 \centering
a)\includegraphics*[angle=-90,width=0.48\textwidth]{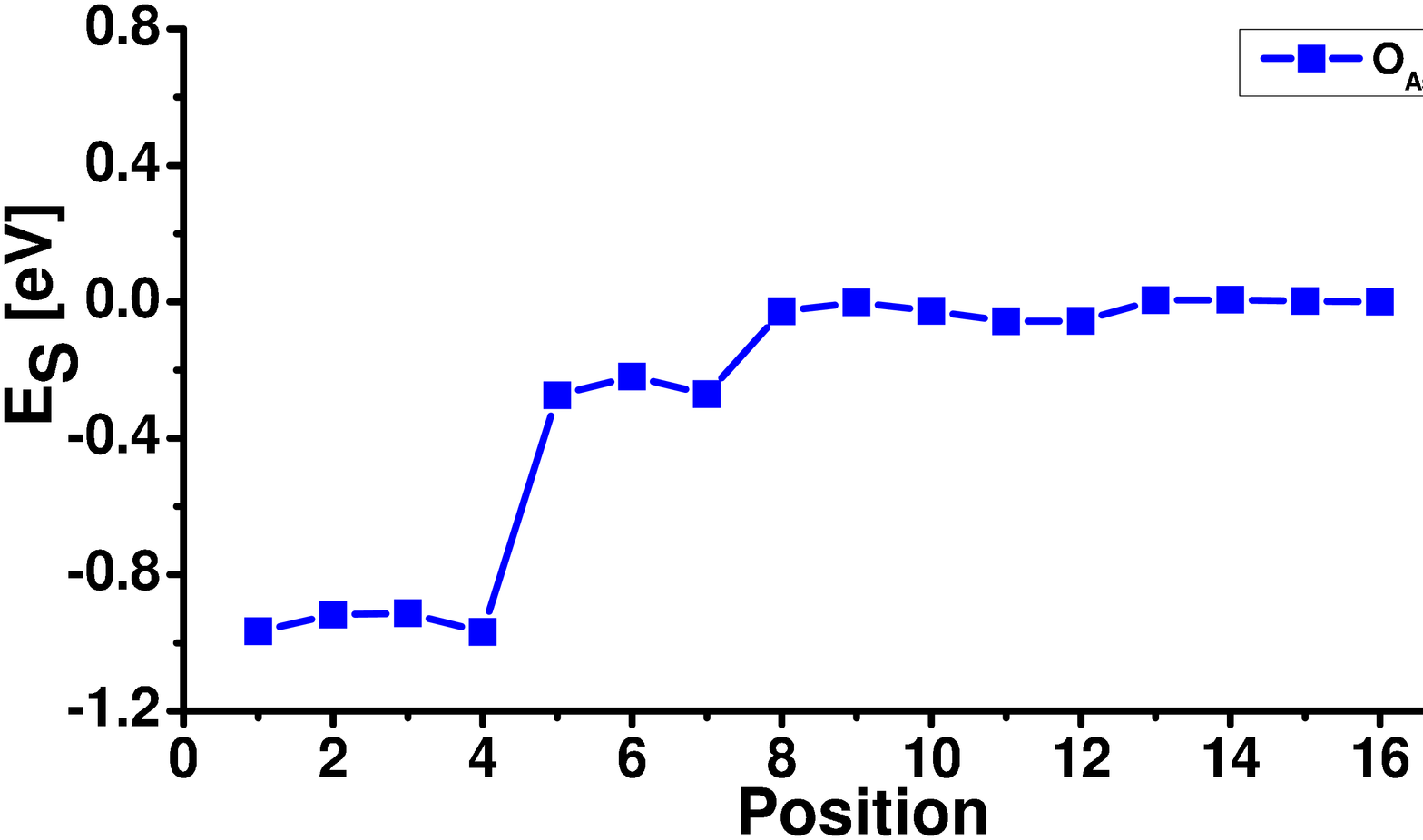}
b)\includegraphics*[angle=-90,width=0.48\textwidth]{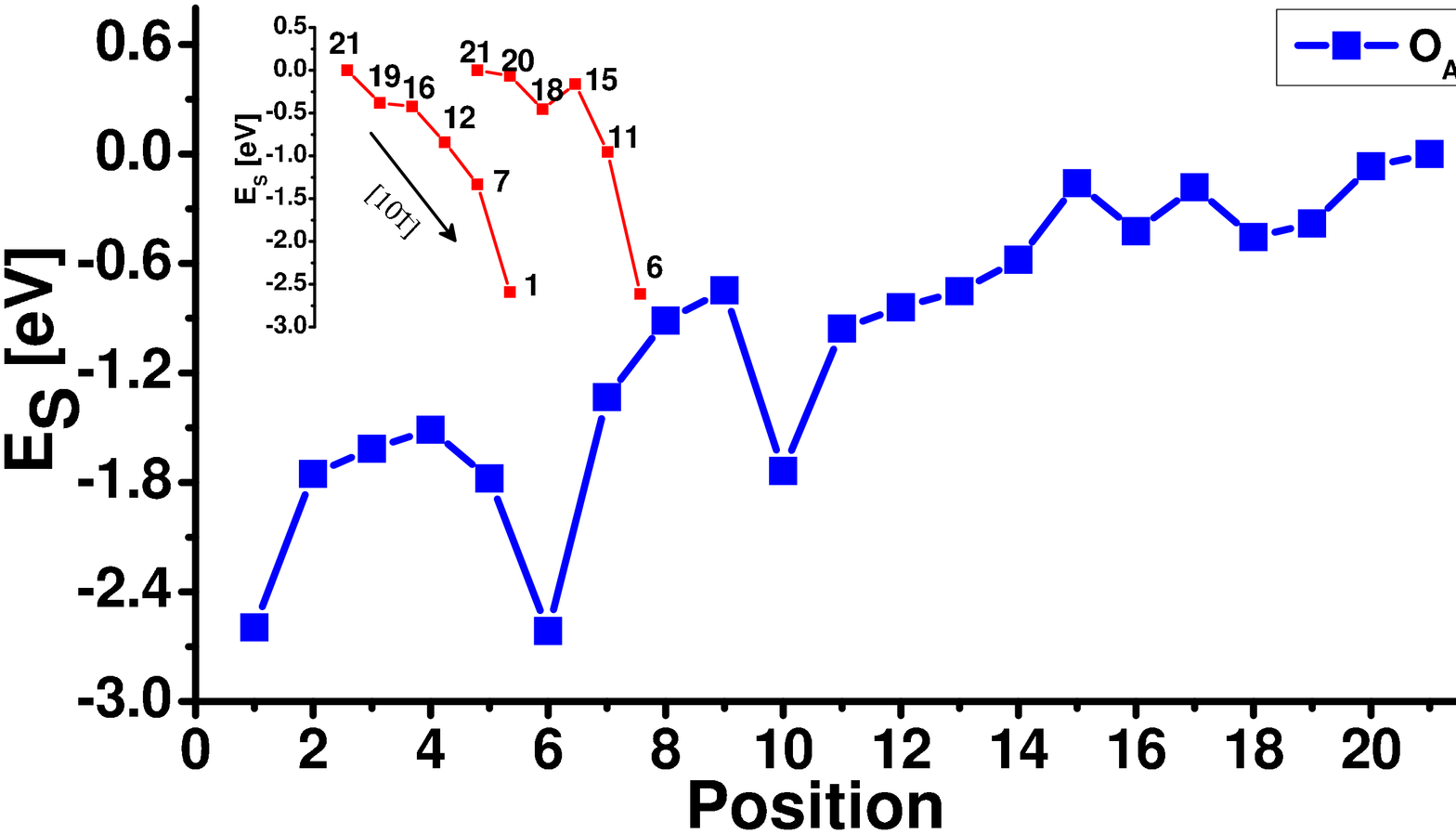}
\caption{(Color online) The segregation energy in a) wz; b) zb GaAs NW as a function of different O atom positions. In the inset to b) the segregation energies for the O atom in subsequent sites along the two nonequivalent $[1,0,\overline{1}]$ lines are shown}
  \label{fig:segregation_o}
\end{figure}

As one can see in Fig.~\ref{fig:segregation_au}, in both wz and zb NWs the Au atoms tend to accumulate near the
lateral surfaces. In wz structure the energy is the lowest for the dopant occupying the threefold coordinated
sites 1 and 4, the corner sites of the sidewall facets. In NWs of zb structure the most energetically favorable
for Au atom is to substitute the cation in position 1, i.e., the cation with extra dangling bond in the corner of
the cross-section of the wire. Analogous results are obtained for oxygen, another element, which can be unintentionally
incorporated during the growth, as shown for GaAs wires in Fig.~\ref{fig:segregation_o}. The difference between the
energy of the wire with an Au or O atom at the lateral surface and at the center of the wire is significant, from
ca 0.6 eV for gold in wz InAs NWs to nearly 3 eV for oxygen in GaAs zb NWs. Thus, both these impurities should indeed
be trapped at the lateral surfaces of the wires. In zb NWs we note that the segregation energy decreases with the
distance from the center of the wire monotonically along the different crystallographic axes, as shown for example
in the inset to Fig.~\ref{fig:segregation_o} for two nonequivalent $[1,0,\overline{1}]$ directions.

\begin{figure}[hbt]
 \centering
a)\includegraphics*[angle=-90,width=0.48\textwidth]{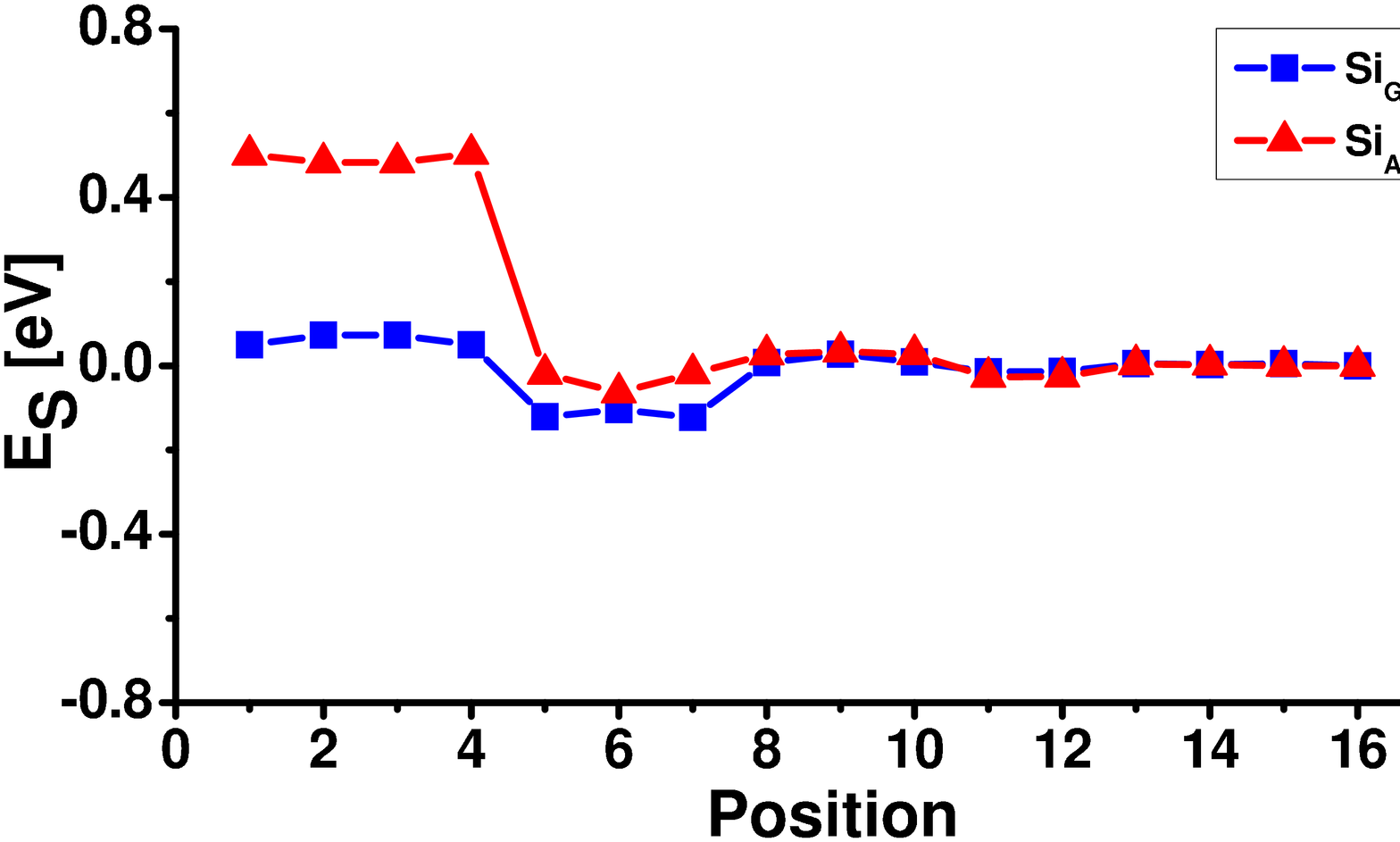}
b)\includegraphics*[angle=-90,width=0.48\textwidth]{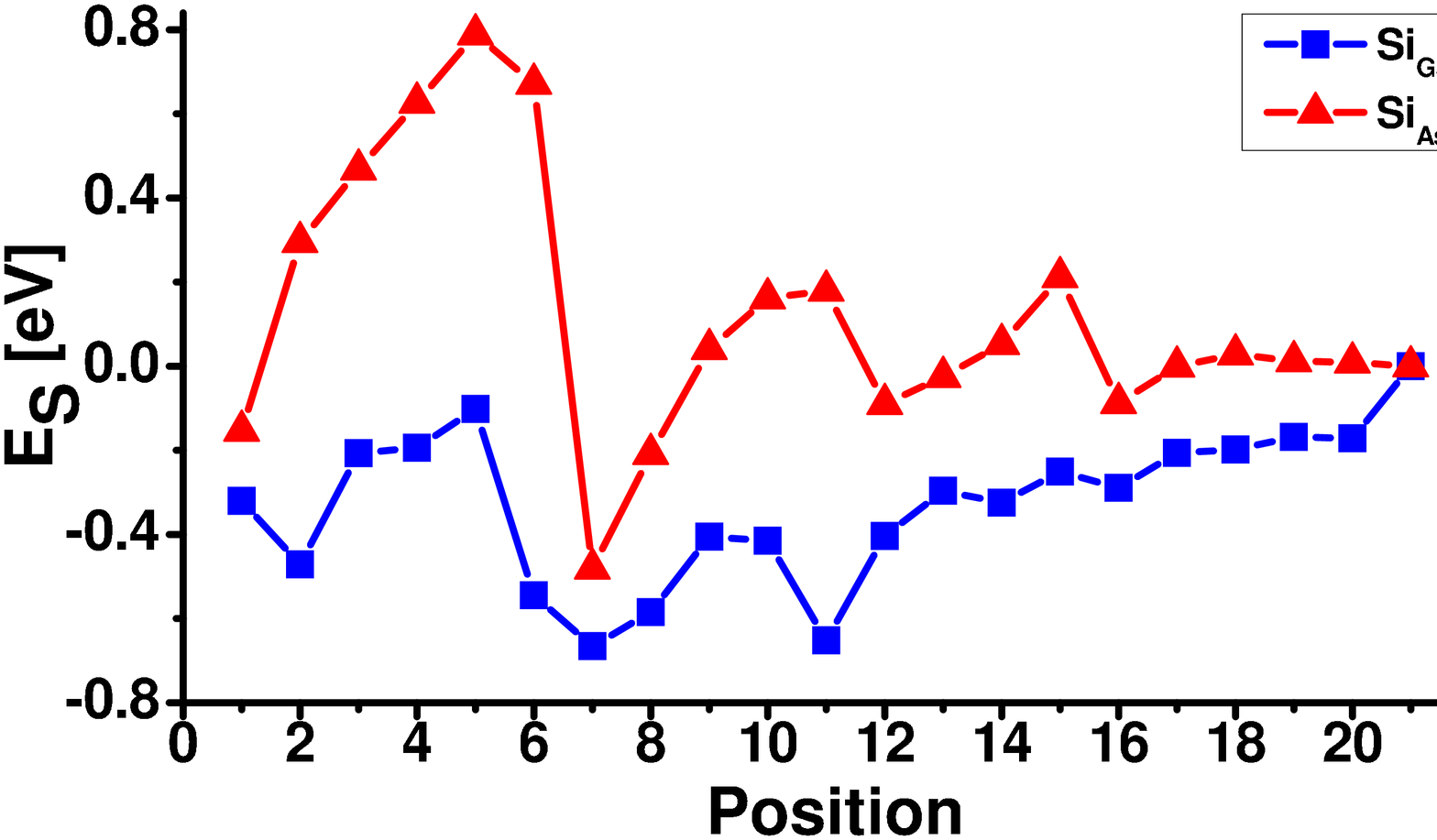}
\caption{(Color online) The segregation energy in a) wz; b) zb GaAs NW as a function of different Si atoms positions.}
  \label{fig:segregation_Si}
\end{figure}

Next, we consider Si doping in GaAs NWs. As already mentioned, since Si is a group IV element in III-V compounds
it can serve either as a donor, by substituting the cation, or as an acceptor, by occupying the anion site. We
therefore analyze the NWs with the Si atom in various cation and anion sites. In Fig.~\ref{fig:segregation_Si}
the results for both wz and zb NWs are presented.
We recall that Si-doped GaAs NWs of wz structure were already studied by Ghaderi {\it{et al.}} in Ref.
\onlinecite{Ghaderi}. These authors considered only four dopant positions in a wz NW, i.e., one center-like,
two surface threefold (distinction is made for corner or middle of facets positions), and one subsurface fourfold
coordinated position. Our results presented in Fig.~\ref{fig:segregation_Si}a, which are obtained for many more Si
positions in a thicker wz GaAs NW, fully confirm the conclusion of Ref. \onlinecite{Ghaderi}. Namely, the four-coordinated
subsurface positions are the most energetically favorable locations for Si acting as a donor as well as an acceptor.
The segregation in these sites is, however, not as strong as for Au or O atoms (ca -0.2~eV). The calculated segregation
energies for Si-doped GaAs NWs of zb structure, which, we recall, were not considered by the authors of Ref.
\onlinecite{Ghaderi}, are presented in Fig.~\ref{fig:segregation_Si}b. As one can notice by comparing Fig.~\ref{fig:segregation_Si}a
and b, the spread of the segregation energy values is considerably larger for zb wires ($\pm 0.7$~eV). It means that
one can expect much more homogeneous distribution of Si atoms in wz wires than in those of zb structure. For both Si
donors and acceptors the most preferable position in zb GaAs wires is site 7, i.e., the fourfold coordinated site next
to site 1. However, for Si substituting Ga (donor) low energy is also obtained for site 11, in the vicinity of the
corner anion with an extra dangling bond. We can thus conclude that our calculations suggest that while in wz GaAs NWs
the energy barrier at the lateral surface is not very high and Si can be incorporated more or less homogeneously, in zb
GaAs NWs Si doping should result in accumulation of Si at the shell.

\begin{figure}[hbt]
 \centering
a)\includegraphics*[angle=-90,width=0.48\textwidth]{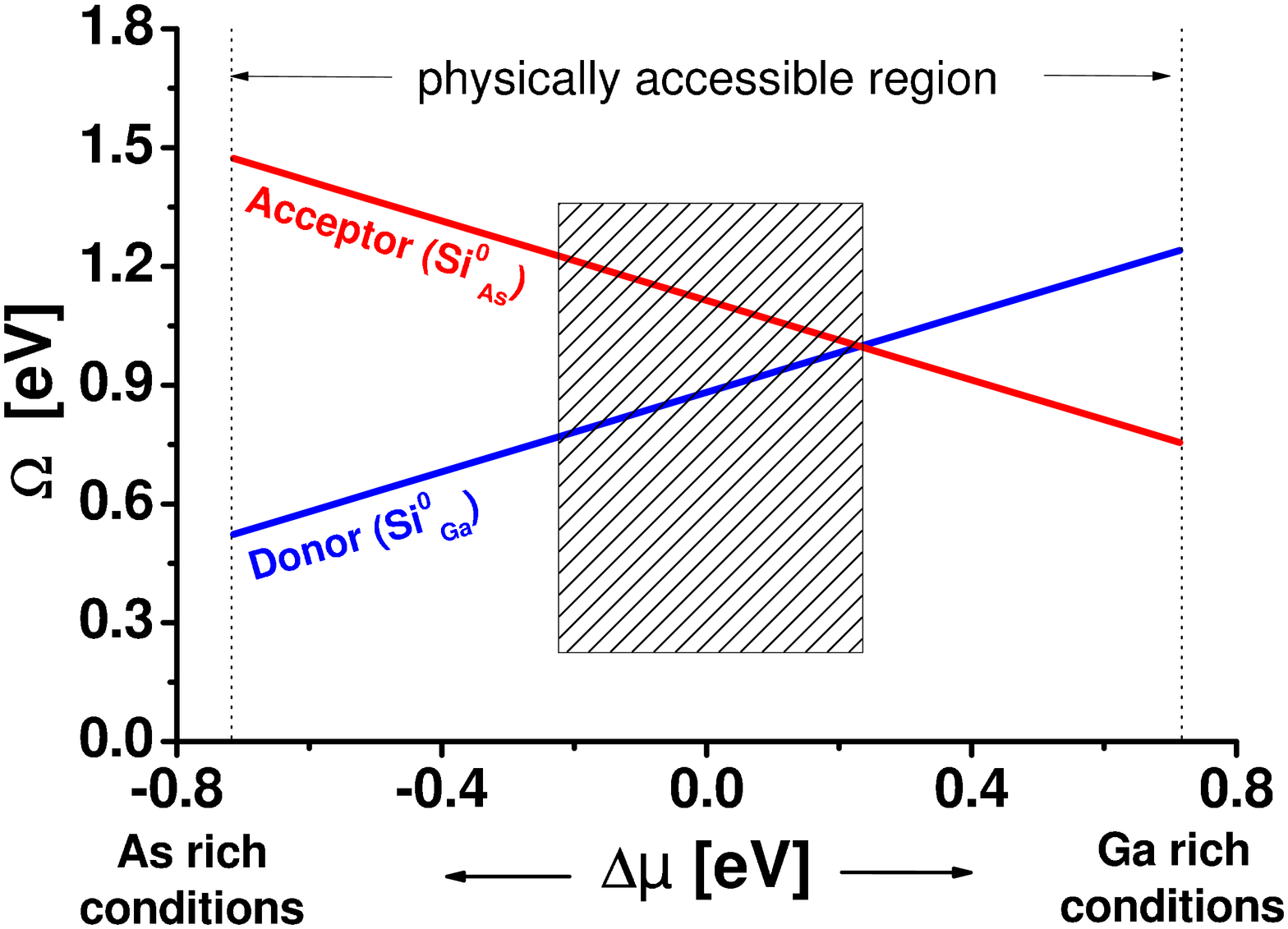}
b)\includegraphics*[angle=-90,width=0.48\textwidth]{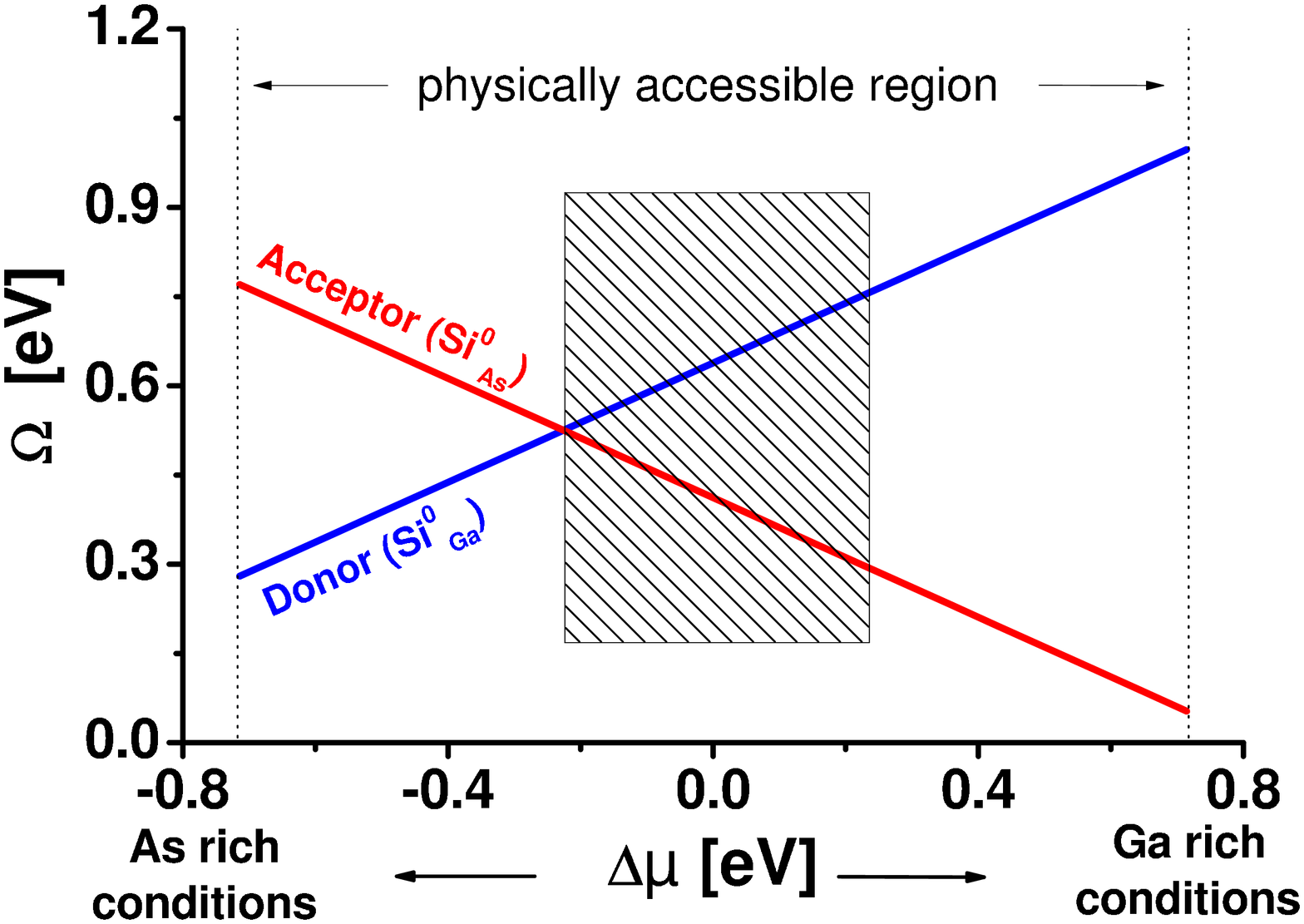}
\caption{(Color online) The formation energy $\Omega$ of GaAs NW with Si impurity  in the center of a) wz NW, b) zb NW as a function of the value $\Delta\mu$.}
  \label{fig:formation_Si}
\end{figure}

The impact of the crystal structure on the incorporation of Si into the III-V NWs is presented in Fig.~\ref{fig:formation_Si}. In this figure the formation energies for Si substituting Ga and As in the center (site 16) of the wz GaAs NW (Fig.~\ref{fig:formation_Si} a) and center (site 21) of zb GaAs NW (Fig.~\ref{fig:formation_Si} b) are compared. We observe that there exists a region of $\Delta\mu$ in which the same growth conditions lead to lower energy for Si substituting Ga in the wire of wz structure and substituting As in the wire of zb structure.

Next, we have calculated the density of states (DOS) for Si-doped GaAs nanowires. In wz structure wires Si acts as expected, i.e., as a donor when substituting Ga and as acceptor when substituting As. Impurity bands are located close to valence band maximum (VBM) and conduction band minimum (CBM) for Si acceptors and donors, respectively. The passivation of the lateral surface have almost no influence on the position of these states, similarly to what was shown for CdTe NWs of wz structure in the paper of T. Sadowski.\cite{TSadowski} In contrast, in not passivated zb NWs we observed surface states in the band gap and the impurity states (for both Si at As and Ga sites) located  at these surface states.
\begin{figure}[hbt]
a)\includegraphics*[angle=-90,width=0.4\textwidth]{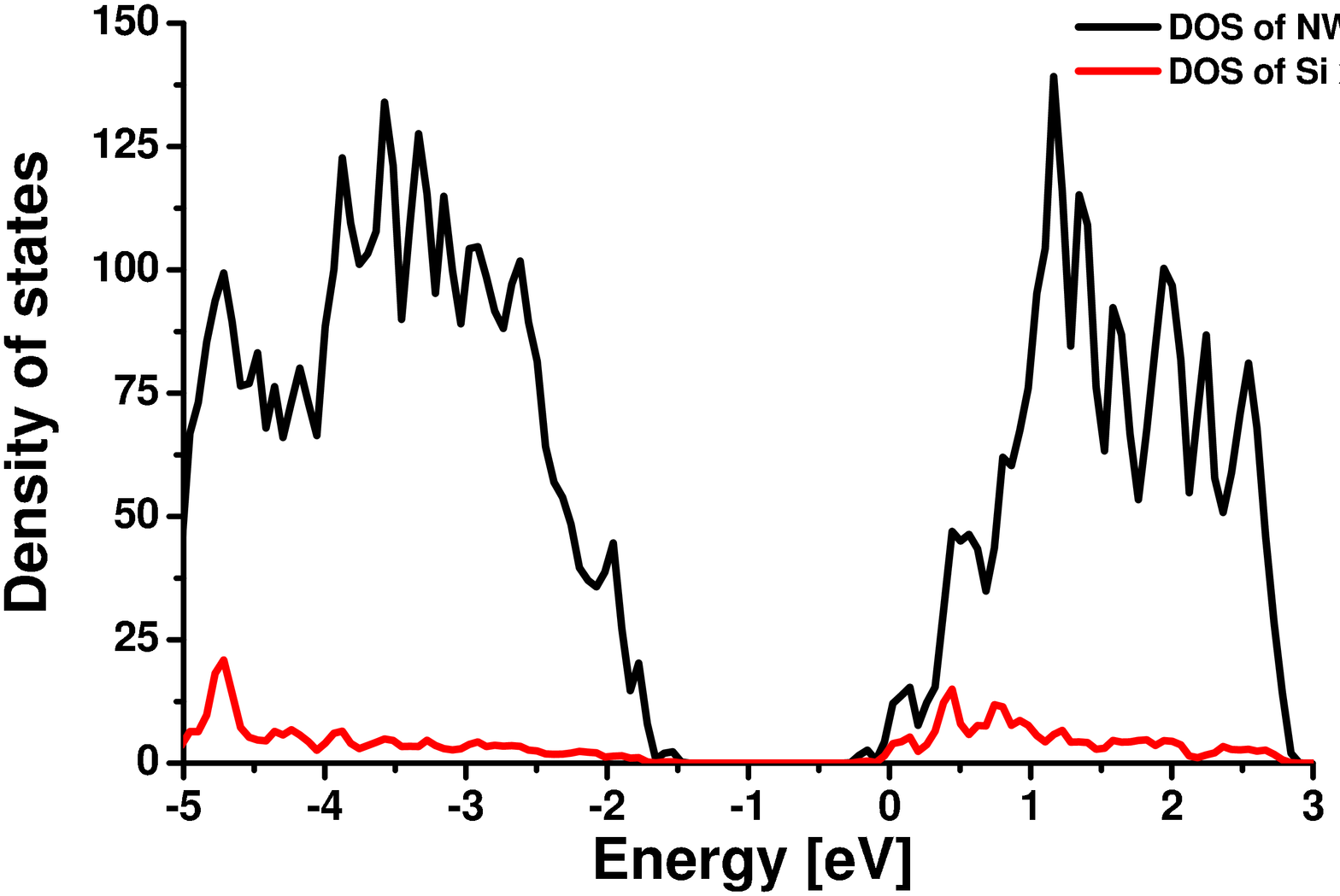}
b)\includegraphics*[angle=-90,width=0.4\textwidth]{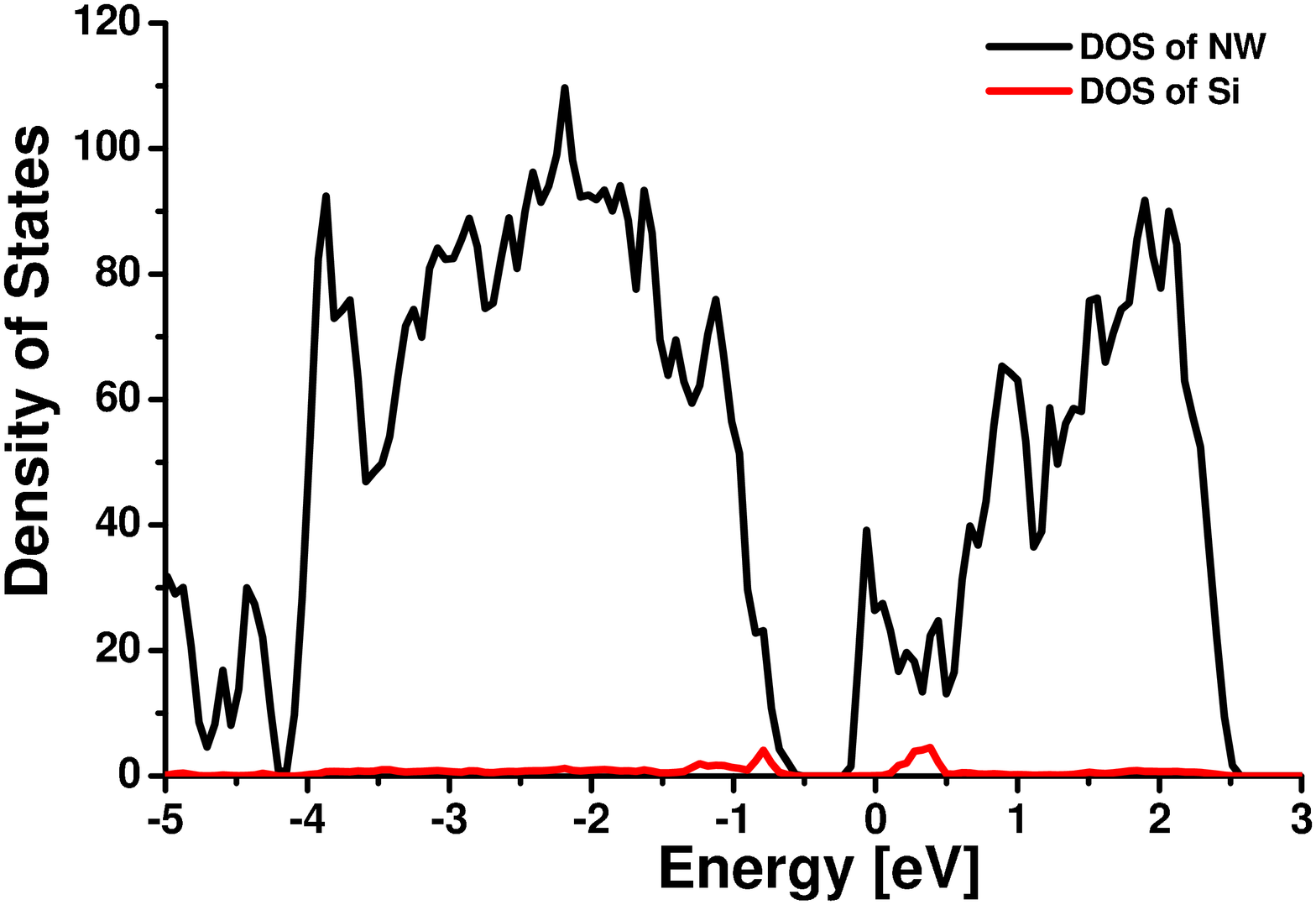}
c)\includegraphics*[angle=-90,width=0.4\textwidth]{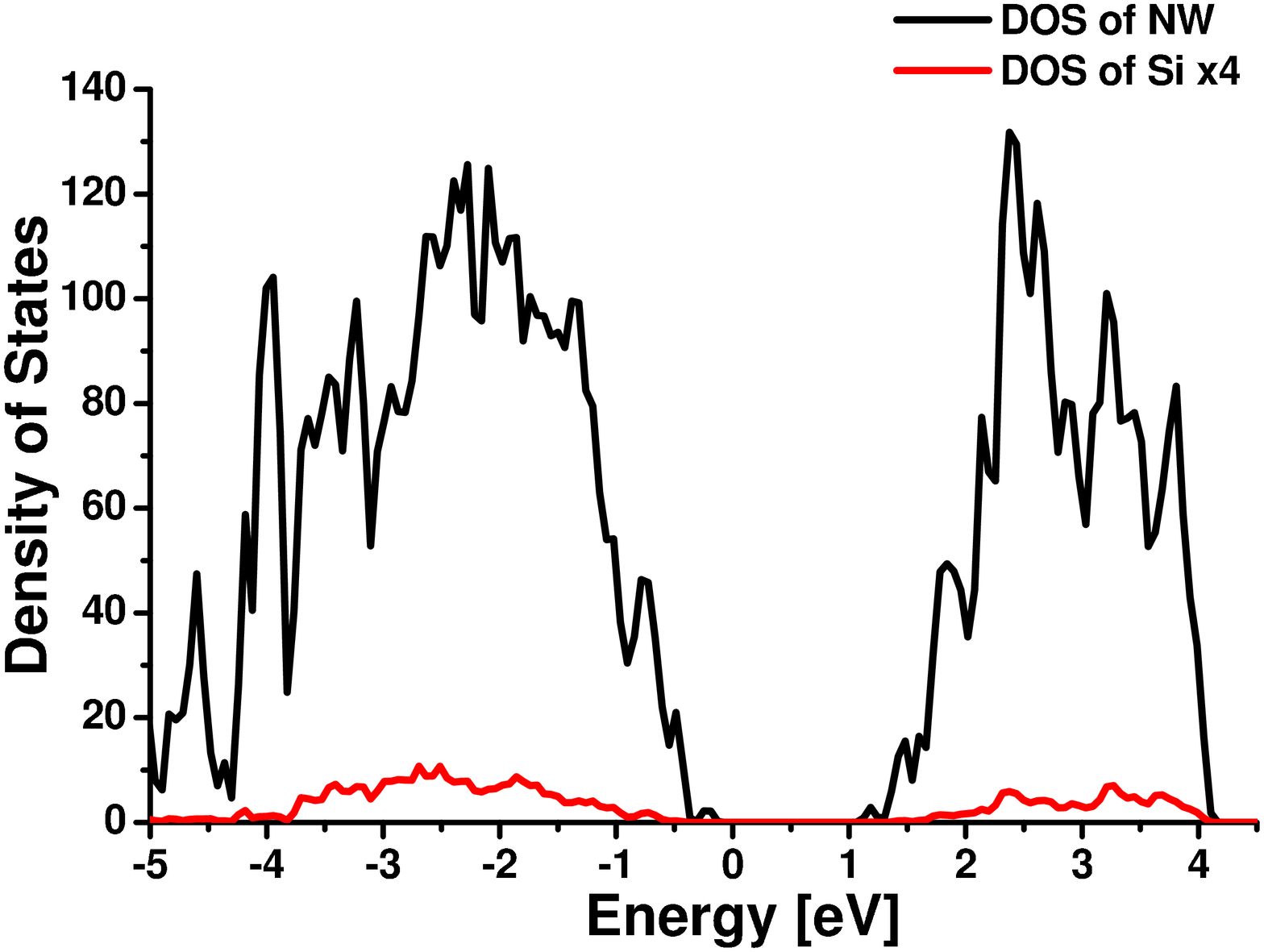}
\caption{(Color online) The density of states calculated for zb GaAs NW with Si substituting Ga atom in the site 21 a) and site 1, c) and b), (site numbers as denoted in Fig.\ref{fig:positions}). The DOS presented in part b) is for the NW with unpassivated lateral surfaces; the results in parts a) and c) are for NWs with lateral surfaces passivated by H atoms. Zero on energy scale denotes always the Fermi level.}
  \label{fig:dos_Si}
\end{figure}
After saturation the surface states disappear and the impurity states move close to VBM and CBM, for Si atom substituting an anion or cation atom, respectively, inside the wire core. In Fig.~\ref{fig:dos_Si}\,a) we show the density of states calculated for H-passivated zb GaAs nanowire with Si substituting the Ga atom in the wire's center position 21. The situation is quite different when Si substitutes the Ga atom in position 1, in the corner of the wire cross-section. For unpassivated nanowires (part b of Fig.~\ref{fig:dos_Si}) Si states and the Fermi level are again pinned at the surface states in the band gap, as for the other impurity positions in zb NWs. However, after passivation  the Fermi level moves to the VBM (compare part c of Fig.~\ref{fig:dos_Si}), what indicates that in this position Si will not behave as an n-type dopant. As the segregation energies suggest that most of the Si atoms are trapped around these corner positions, this behavior can explain the observed dominance of p-type in Si-doped GaAs NWs.~\cite{Piccin}

\begin{figure}[hbt]
a)\includegraphics*[angle=-90,width=0.48\textwidth]{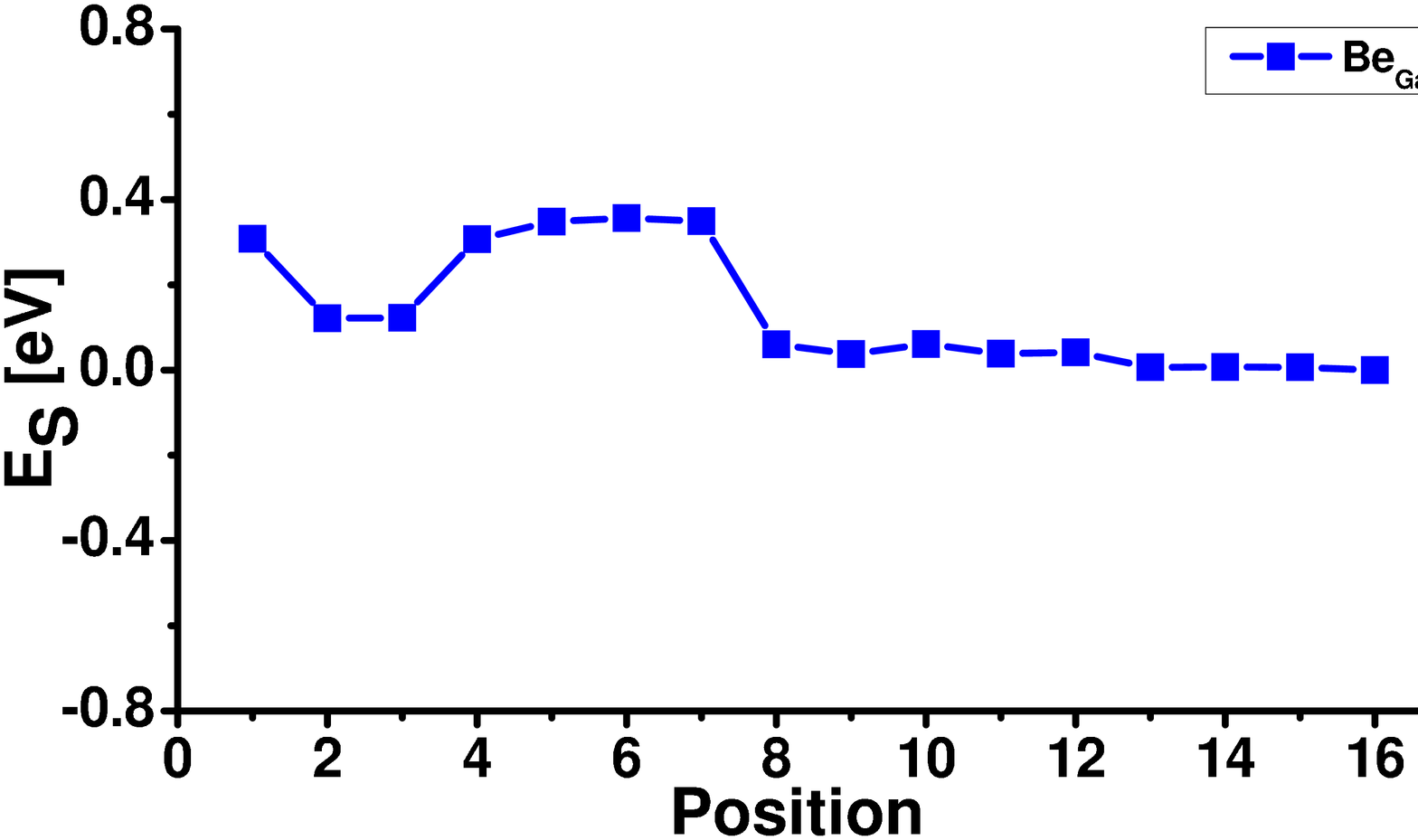}
b)\includegraphics*[angle=-90,width=0.48\textwidth]{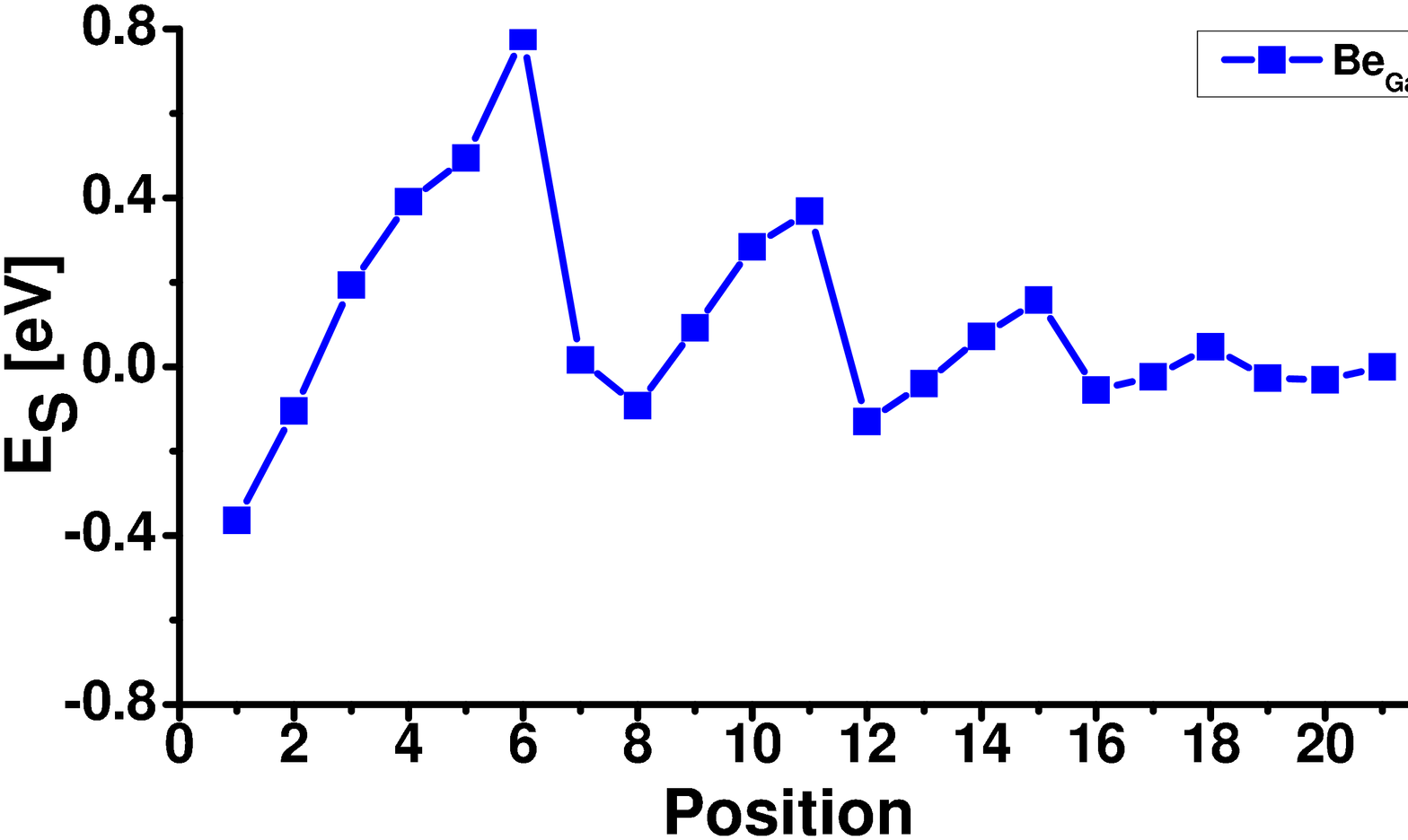}
\caption{(Color online) The segregation energy in a) wz; b) zb GaAs NW as a function of different Be atom positions.}
  \label{fig:segregation_Be}
\end{figure}

In the next step we compare the segregation energy obtained for Si-doped GaAs NWs with the segregation energy in wires doped with beryllium,
one of the most widely used acceptors in GaAs. Results obtained for Be-doped GaAs NW of wz and zb structure are presented in Fig.~\ref{fig:segregation_Be}.
As shown in the Figure, in wz NWs the distribution of Be atoms should be again quite homogenous, but in this case
the lowest energy corresponds to the situation where the impurity substitutes an atom inside the wire. Again, in zb
GaAs NWs the lowest value of the segregation energy (-0.4~eV) was obtained when Be substitutes the Ga atom with an
extra dangling bond in site 1. It should be noted, however, that this is much weaker segregation than obtained in
zb GaAs NWs for Si, and of course Au and O.
Therefore, these results seem to be in agreement with the experimental
observation that even though Be atoms accumulate in the shell of zb GaAs NWs, it should be possible to dope the NWs
almost homogeneously during axial growth.\cite{casadei} It is also interesting to note that for some reason the lowest
energies of the zb GaAs wires with Be atoms are found when the impurities are located in sites along the $[1,0,\overline{1}]$
axis (from the center of the wire to the cation with extra dangling bond at the corner of the lateral surface). This
is in contrast to the situation presented in Fig.~\ref{fig:segregation_o} for oxygen and gold impurities. The low energy
cost of incorporation of Be along the $[1,0,\overline{1}]$ axis can explain the recently observed diffusion of this
impurity into the volume of the zb GaAs NWs, interpreted in Ref. \onlinecite{casadei} as an alternative incorporation path.

\begin{figure}[hbt]
 \centering
\includegraphics*[angle=-90,width=0.48\textwidth]{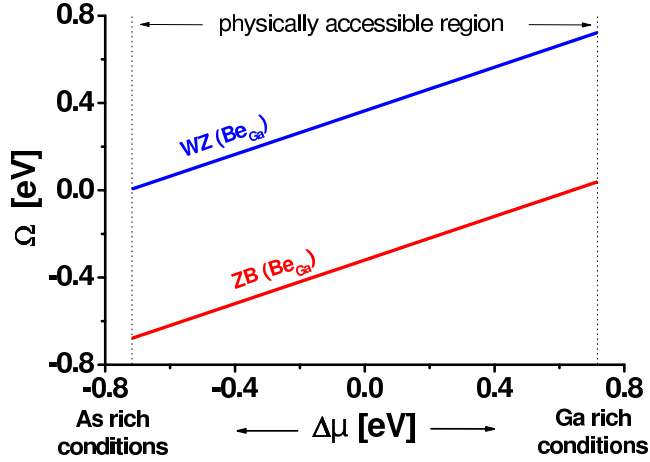}
\caption{(Color online) The formation energy $\Omega$ of zb and wz GaAs NW with Be impurity as a function of the value $\Delta\mu$.}
  \label{fig:formation_Be}
\end{figure}

Finally, the comparison of formation energies for Be incorporation into the GaAs NWs, presented in Fig.~\ref{fig:formation_Be}, shows that it should be much easier to dope with beryllium the NWs with zb structure than those with wz crystal structure. It shows also that the formation energies for Be are much lower {\bf\color{red}than} for Si.

\begin{figure}[!hbt]
a)\includegraphics*[angle=-90,width=0.48\textwidth]{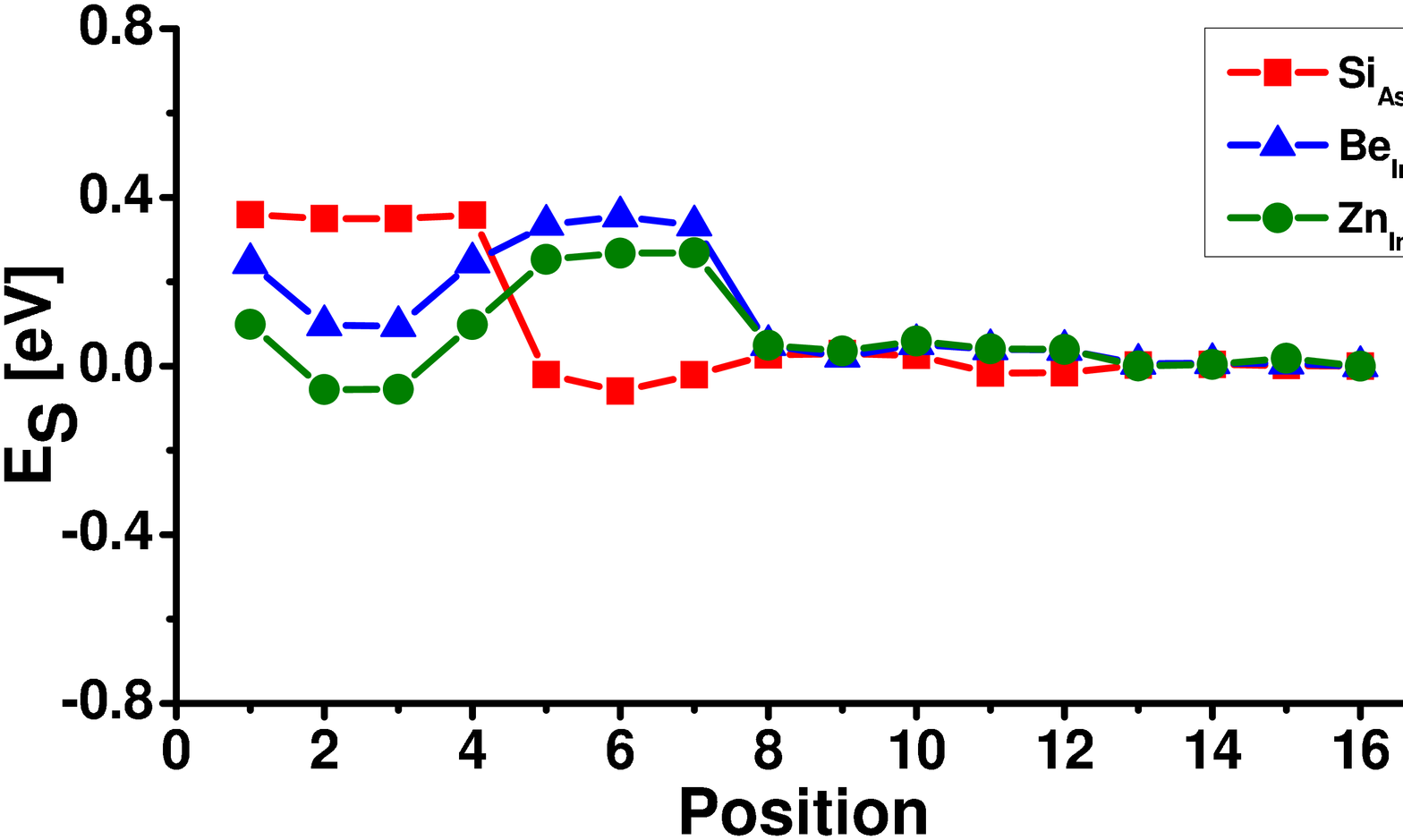}
b)\includegraphics*[angle=-90,width=0.48\textwidth]{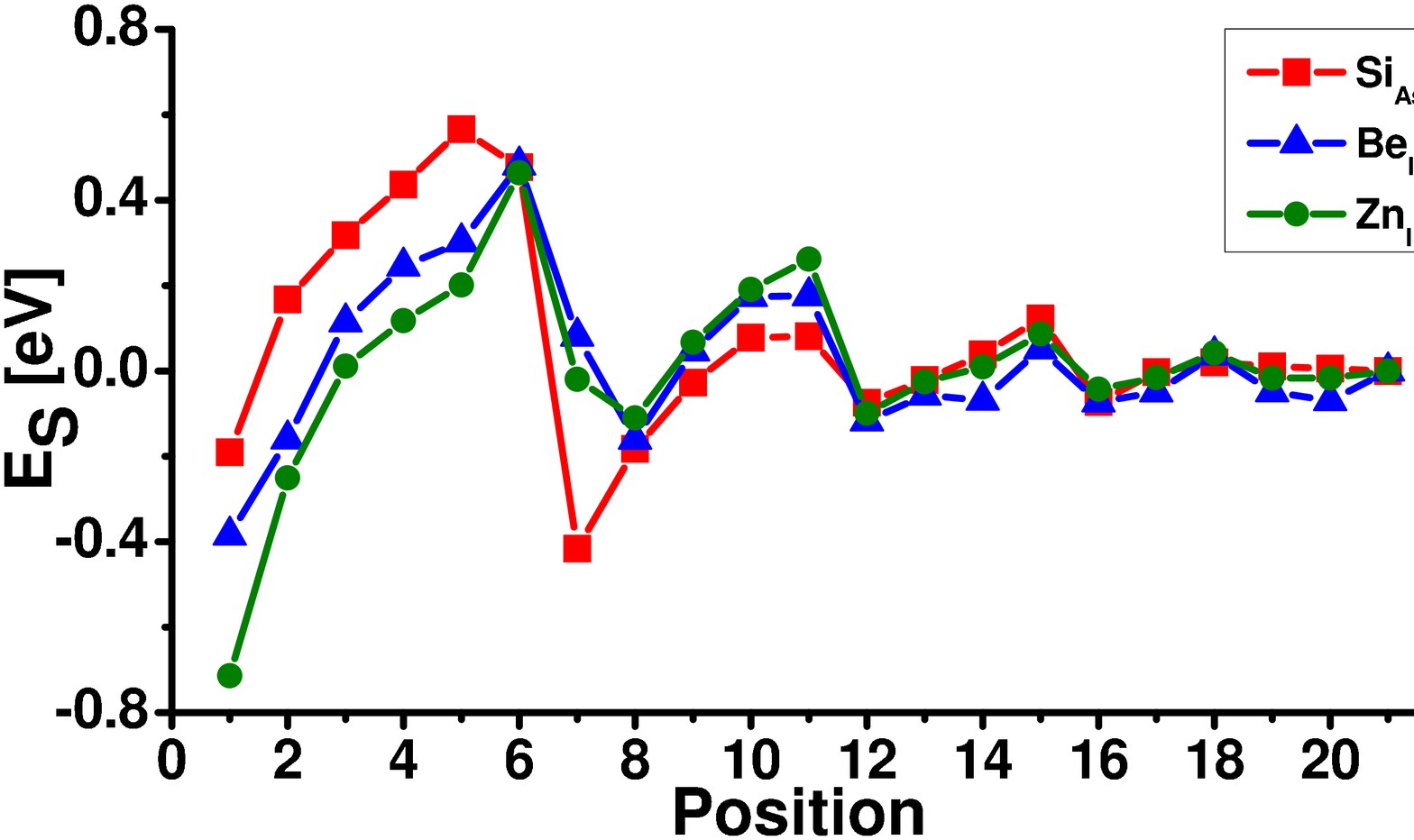}
\caption{(Color online) The segregation energy for Si, Be and Zn acceptors in various positions in wz (a) and zb (b) InAs NW.}
  \label{fig:segregation_inas}
\end{figure}

\begin{figure}[!hbt]
\includegraphics*[width=0.45\textwidth]{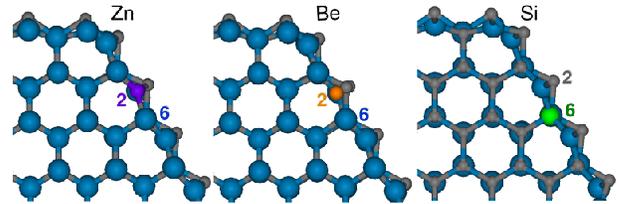}
\caption{(Color online) The most energetically favorable positions of Zn, Be and Si acceptors at the lateral surfaces of wz InAs NW.}
  \label{fig:positions_inas}
\end{figure}

As mentioned before, the as grown InAs NWs are usually n-type. Doping with several acceptors, like Be or Zn, have been
tried to obtain p-type conductivity in these NWs.\cite{Sorensen, Ford} In Fig.~\ref{fig:segregation_inas}
we present the comparison of segregation energies obtained for InAs NWs doped with Si, Be and Zn atoms as a function of
their different locations. The study of segregation energy for different acceptors in InAs NWs shows that in these wires,
similarly to GaAs NWs, wz structure should lead to a more homogeneous distribution of impurities. Inside the wire, in the
core, all impurity positions are energetically equivalent in both structures. In the shell, Be and Zn prefer to be in the
middle of the sidewall of the wire, substituting threefold coordinated cations, while Si, substitutes fourfold coordinated
anion at the NW's lateral surface (site 6), as presented in Fig.~\ref{fig:positions_inas}. One can understand this difference
by recalling that Zn and Be occupy the cation positions while Si, should substitute an anion to serve as an acceptor. The
relaxation of the atom positions at the wire lateral surfaces leads to different reconstruction of bonds for cations and
anions, as shown in Ref. \onlinecite{JPCM} for pure III-V NWs. The most important result of this study is, however, the
fact that in InAs NWs of wz structure the segregation energy for Be atoms is
not negative for any of the considered positions. Also for Si in wz InAs NWs the segregation energy for the impurity in
site 6, although negative, is close to zero (-0.06~eV). Thus, we can conclude that the best precursor for p-type doping
of InAs NWs should be beryllium but using Si can also be effective, provided the wz crystal structure of the wires is assured.

\begin{figure}[!hbt]
a)\includegraphics*[angle=-90,width=0.48\textwidth]{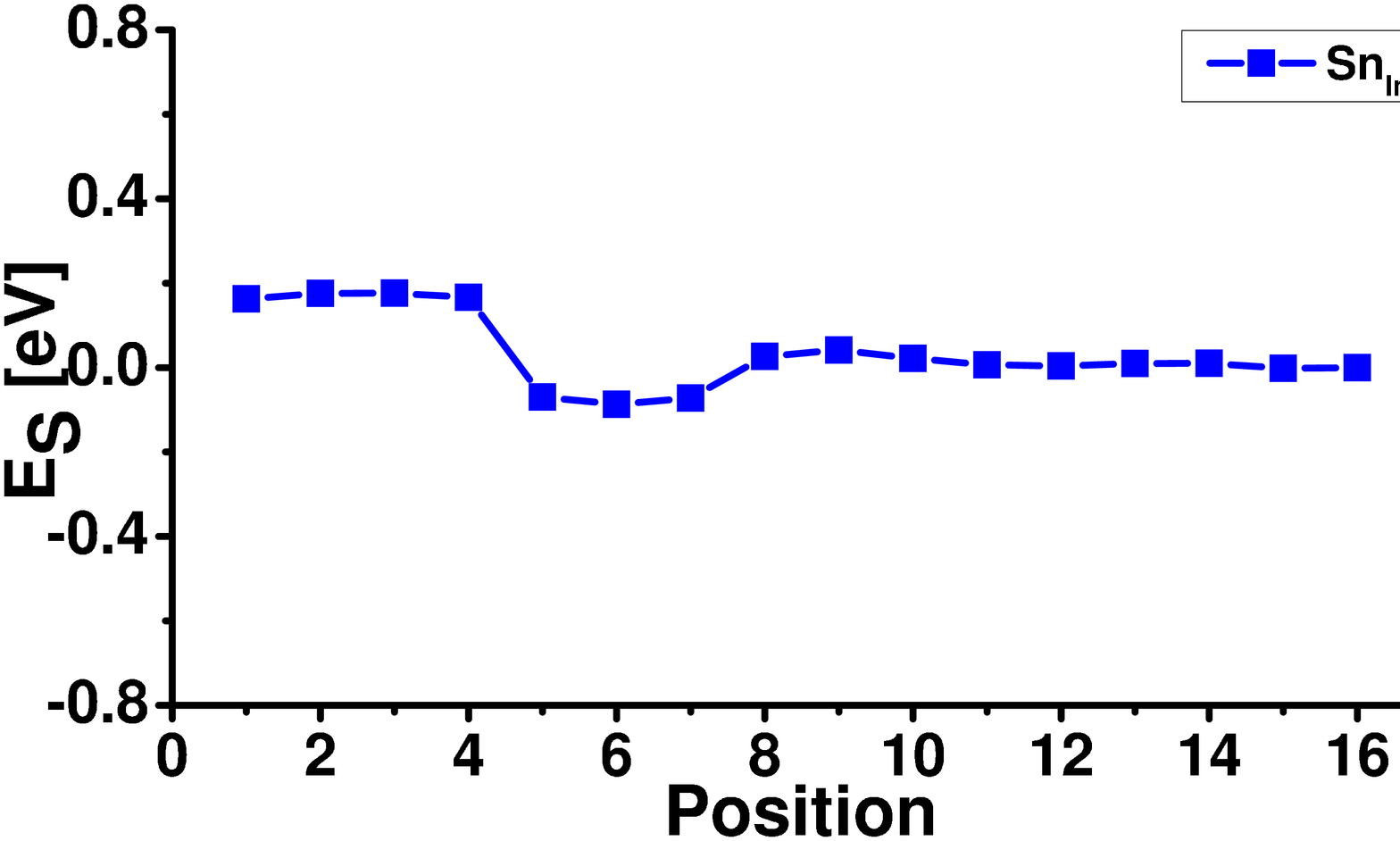}
b)\includegraphics*[angle=-90,width=0.48\textwidth]{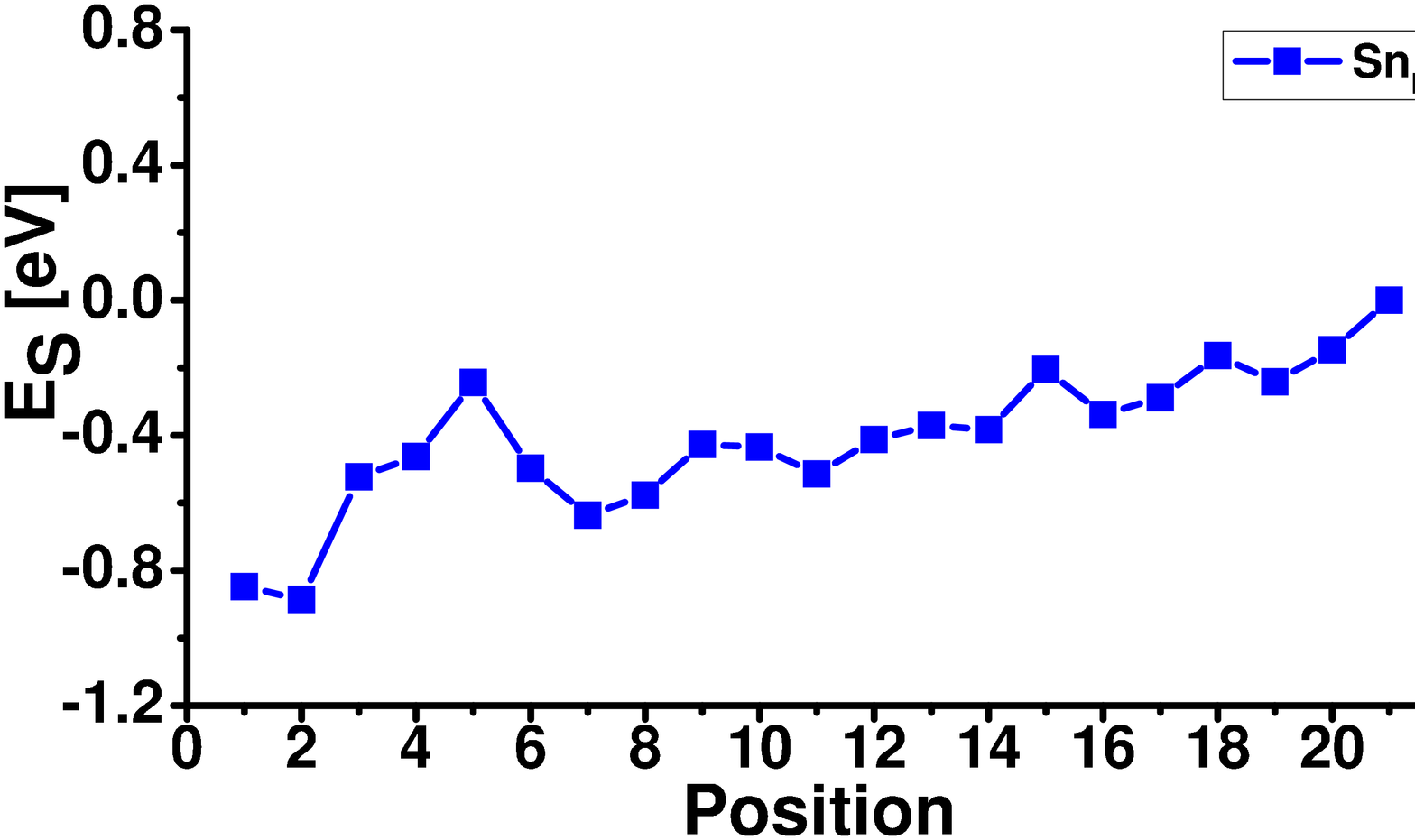}
\caption{(Color online) The segregation energy of a) wz; b) zb InAs NW as a function of different Sn positions.}
  \label{fig:segregation_inas_sn}
\end{figure}
In Ref. \onlinecite{Astromskas} it was shown, by capacitance measurements, that the concentration of carriers in InAs
NWs can be increased by n-type doping with Sn. In Fig.~\ref{fig:segregation_inas_sn} the segregation energy obtained for
InAs NWs doped with Sn is presented. The results for Sn in wz InAs wires are very similar to that for Si-doped InAs NWs,
presented above. Tin ions distribute fairly homogeneously across the wz wire. Although for Sn the most favorable is to
substitute the fourfold coordinated cations at the wz NW's lateral surface, the segregation energy is again very small.
This result seems to agree with the observed increase of the carrier concentration and the surface charge density in
Sn-doped wz InAs NWs.\cite{Astromskas} It should be noted that in Ref. \onlinecite{ThelanderNanotech} it was shown that
the Sn precursor increases the stacking fault density in wz NWs, ultimately at high flows leading to a zb crystal structure
with strong overgrowth and very low resistivity. In agreement with this observation we obtain that in zb NWs, analogously to
other dopants in this material, Sn atoms do not incorporate into the wire but stay at the lateral surface, with particular
reference to atoms with extra dangling bond (Fig.~\ref{fig:segregation_inas_sn}b).

\begin{figure}[hbt]
 \centering
a)\includegraphics*[angle=-90,width=0.48\textwidth]{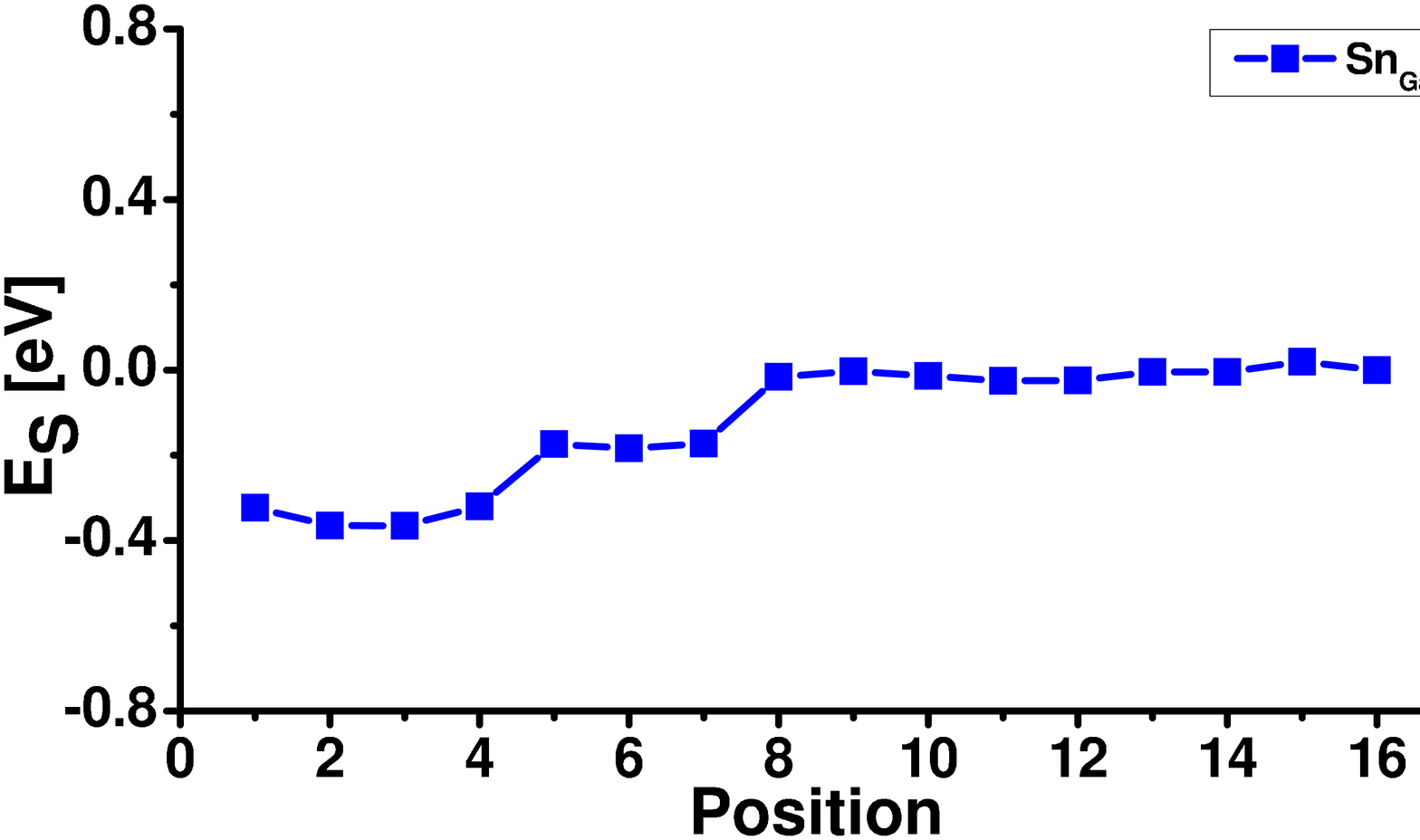}
b)\includegraphics*[angle=-90,width=0.48\textwidth]{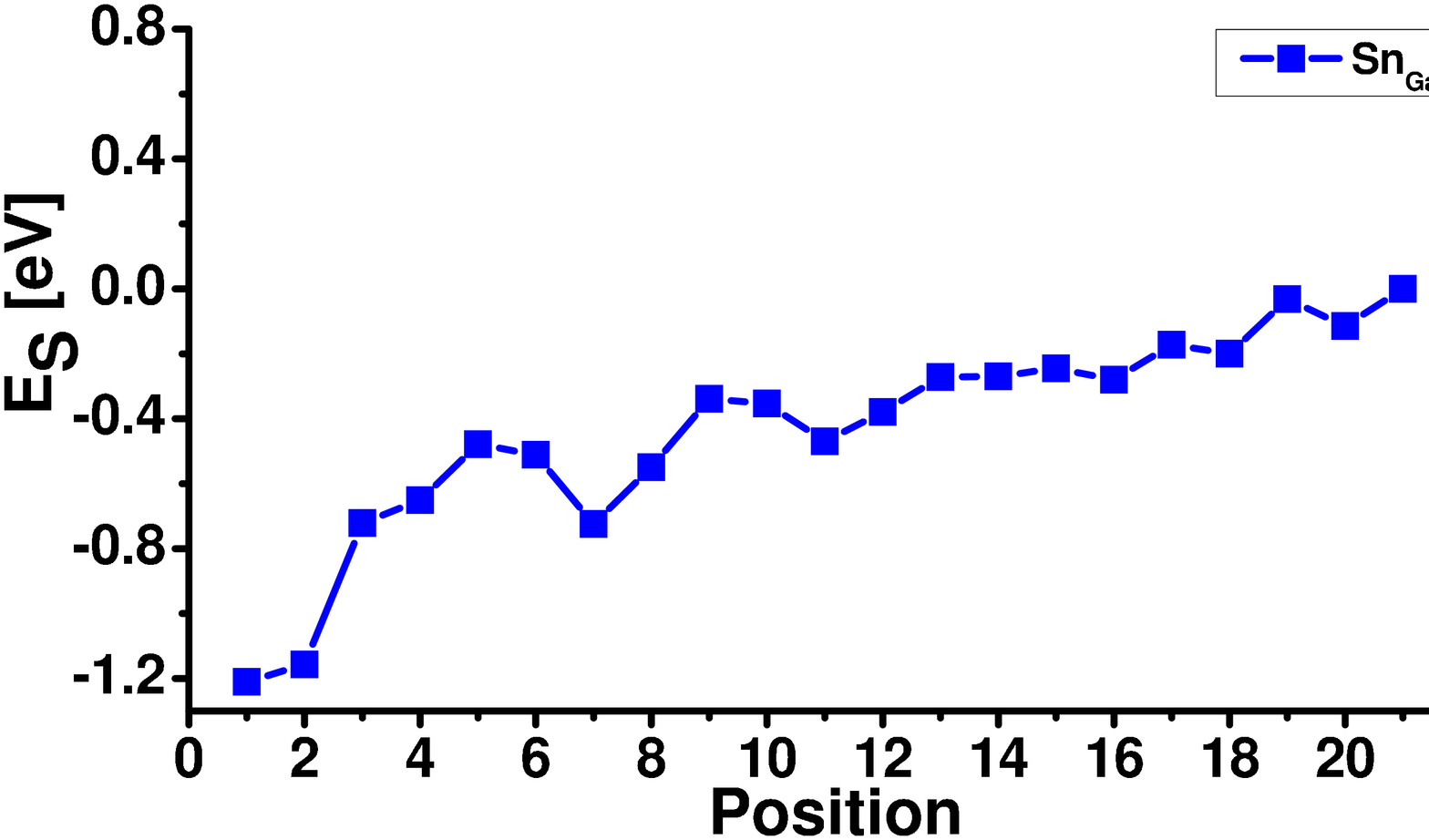}
\caption{(Color online) The segregation energy of a) wz; b) zb GaAs NW as a function of different Sn atoms positions.}
  \label{fig:segregation_sn}
\end{figure}

In contrast, our calculations performed for GaAs NWs show that in this material Sn atoms tend to segregate to the lateral
surfaces in both wz and zb structures (see Fig.~\ref{fig:segregation_sn}). Thus, although our calculations performed for Sn
impurities in III-V wires agree with the observation
that doping with tin can be used to increase the concentration of electrons in InAs wires,\cite{Astromskas}
they suggest that using Sn for n-type doping of GaAs NWs can be much less efficient.

\section{Conclusions}
Nanowire growth processes usually rely on the vapor - liquid - solid (VLS) mechanism normally assisted by gold droplets
or self assisted on a SiO/SiO$_2$ surface. Doping these structures during such complex growth process is difficult to
understand and control. It is, however, already widely accepted that there are two distinct pathways for dopant
incorporation  into semiconductor NWs: the VLS mechanism, in which the dopant atoms first dissolve in
the catalyst and are then transported across the liquid –- solid interface to the core of the NW, and
the vapor –- solid (VS) mechanism, when dopant atoms are directly deposited on the lateral NW surfaces.
In Ref. \onlinecite{Lauhon} it has been suggested that these two mechanisms proceed at different rates --
namely, the latter predominates and leads to a much higher concentration of dopants in the shell of the
NW than in the core. This conclusion was based on the experimentally observed accumulation of gold atoms
close to the lateral surfaces of InAs NWs\,\cite{Perea} as well as phosphorus in Ge
NWs.\cite{Lauhon} Radovanovic and co-workers, when studying GaN wires doped with Co and Cr, have also
concluded that adsorption on the lateral surface is the main doping mechanism of the semiconductor
NWs.\cite{Radovanovic} Finally, the conclusion that Be atoms are mostly incorporated into the zb GaAs
NWs from the side facets and that the incorporation through the Ga droplet is negligible are made in
Ref. \onlinecite{casadei}. It should be also mentioned that the very recently reported beautiful model,\cite{Chen} 
in which B-doped Si NWs were considered, has shown that 2D diffusivity along the solid-liquid NW interface plays 
also a considerable role in impurity incorporation. In this paper we show, by means of first-principles calculations of the total
energy of doped GaAs and InAs NWs, that indeed most of the commonly used impurities for p- and n-type doping
should accumulate near the surfaces of the NW. It should be, however, emphasized that this result does not
depend on the growth mechanism, because in our calculations only the differences in formation energies for
various dopant positions have been taken into account. Thus, it shows that even for incorporation of the
dopants into the core, via VLS mechanism, the segregation of impurities to the surfaces should take place.
The only exceptions are wz GaAs as well as wz InAs NWs doped with Be -- in these the segregation energy is
positive for all studied cation sites substituted by Be atoms.

We note that the degree of segregation depends considerably
on the NW's host material, crystal structure of the wire and of course the particular dopant. Our calculations show
that in NWs of zb structure the segregation for the most common dopants is much higher
than in NWs of wz structure. The studied impurities in zb NWs should remain either at the lateral surfaces,
where they prefer to substitute the atom with extra dangling bond, or should be trapped at the subsurface, in the
vicinity of the atom with additional dangling bond. At these positions the segregation energies in zb GaAs
NWs are lower than in the core by ca 0.4~eV for beryllium, but by even 1.5-3 eV for gold and oxygen. It is also shown that at these positions Si atoms do not act as donors even when substituting Ga cations - this result is in agreement with the observed predominance of p-type in Si-doped GaAs NWs. For gold
and oxygen in wz wires the results are similar to those obtained for zb structure but with lower, though still
high, segregation. With our calculations we have thus confirmed the experimentally observed segregation
of Au atoms to lateral surfaces.\cite{Perea} According to our results the same should be also observed for
oxygen. In contrast, all other studied impurities distribute much more homogeneously across the wz wire and
Be even prefers to substitute the Ga(In) ion at the center of the wire, as mentioned above. We can thus
conclude that our calculations suggest that growth conditions leading to wz structure can help to avoid accumulation of impurity atoms at the surface during the growth of doped III-V wires. It seems
that the best choice for effective p-type doping of both GaAs and InAs wz NWs is to use beryllium as a dopant, due to it's low segregation. However, the cost of substituting the host cation by Be atom, i.e., the formation energy, is lower in the zb than in the wz NWs.
Finally, we observe that the different from other impurities behavior of Au and O can be attributed to the large electronegativity
of these elements (3.5 for O and 2.4 for Au) as compared to electronegativities of Ga, In and all other studied
dopants, which are in the range of 1.5 -- 1.8.

As shown above, our results suggest that the energy of a doped NW can be very different when the impurity position inside the wire changes, especially in zb structure.  Only in the case of oxygen and beryllium in GaAs NWs some regularities in these differences can be observed. In other cases the fluctuations of the segregation energy values seem to be random. These results were fully confirmed by several tests, i.e., by repeating the calculations for the impurities in equivalent sites, by changing the NW diameter and finally by moving slightly the initial atomic positions. Recently, thanks to laser assisted atom probe tomography the studies of the distribution of impurities in semiconductor wires have become possible. Our results may suggest that the observed inhomogeneities (compare, e.g., Ref.~\onlinecite{Chen}) may have a more fundamental origin than just disorder. Neither the noticed trends nor the other segregation fluctuations, which we are not able to explain in our approach, were reported before. It should be noted, however, that in all previous studies the segregation and/or formation energies in doped NWs have been calculated for at most 3-4 different impurity positions. In our study the results for wz NWs with the impurities in 16 different sites and for zb NWs in 21 different sites have been compared - to our best knowledge, most of the nonequivalent sites considered here have not been studied before for any impurity.

\begin{acknowledgments}
We thank Hadas Shtrikman, Peter Krogstrup, Hanka Przybylinska and Ewa Bialkowska-Jaworska for fruitful discussions and valuable comments.
The research leading to these results has received funding from the European Community's 7th Framework Programme
[FP7/2007-2013] under grant agreement n\textdegree 215368, from Foundation for Polish Science and Ministry of
Higher Education (Poland) Grant IP 2011013671. All computations were carried out in the Informatics Center
Tricity Academic Computer Net (CI TASK) in Gdansk.
\end{acknowledgments}

\bibliographystyle{apsrev4-1}
\bibliography{referencje}

\begin{thebibliography}{45}%
\makeatletter
\providecommand \@ifxundefined [1]{%
 \@ifx{#1\undefined}
}%
\providecommand \@ifnum [1]{%
 \ifnum #1\expandafter \@firstoftwo
 \else \expandafter \@secondoftwo
 \fi
}%
\providecommand \@ifx [1]{%
 \ifx #1\expandafter \@firstoftwo
 \else \expandafter \@secondoftwo
 \fi
}%
\providecommand \natexlab [1]{#1}%
\providecommand \enquote  [1]{``#1''}%
\providecommand \bibnamefont  [1]{#1}%
\providecommand \bibfnamefont [1]{#1}%
\providecommand \citenamefont [1]{#1}%
\providecommand \href@noop [0]{\@secondoftwo}%
\providecommand \href [0]{\begingroup \@sanitize@url \@href}%
\providecommand \@href[1]{\@@startlink{#1}\@@href}%
\providecommand \@@href[1]{\endgroup#1\@@endlink}%
\providecommand \@sanitize@url [0]{\catcode `\\12\catcode `\$12\catcode
  `\&12\catcode `\#12\catcode `\^12\catcode `\_12\catcode `\%12\relax}%
\providecommand \@@startlink[1]{}%
\providecommand \@@endlink[0]{}%
\providecommand \url  [0]{\begingroup\@sanitize@url \@url }%
\providecommand \@url [1]{\endgroup\@href {#1}{\urlprefix }}%
\providecommand \urlprefix  [0]{URL }%
\providecommand \Eprint [0]{\href }%
\providecommand \doibase [0]{http://dx.doi.org/}%
\providecommand \selectlanguage [0]{\@gobble}%
\providecommand \bibinfo  [0]{\@secondoftwo}%
\providecommand \bibfield  [0]{\@secondoftwo}%
\providecommand \translation [1]{[#1]}%
\providecommand \BibitemOpen [0]{}%
\providecommand \bibitemStop [0]{}%
\providecommand \bibitemNoStop [0]{.\EOS\space}%
\providecommand \EOS [0]{\spacefactor3000\relax}%
\providecommand \BibitemShut  [1]{\csname bibitem#1\endcsname}%
\let\auto@bib@innerbib\@empty
\bibitem [{\citenamefont {Piccin}\ \emph {et~al.}(2007)\citenamefont {Piccin},
  \citenamefont {Bais}, \citenamefont {Grillo}, \citenamefont {Jabeen},
  \citenamefont {Franceschi}, \citenamefont {Carlino}, \citenamefont
  {Lazzarino}, \citenamefont {Romanato}, \citenamefont {Businaro},
  \citenamefont {F.}, \citenamefont {Martelli},\ and\ \citenamefont
  {Franciosi}}]{Piccin}%
  \BibitemOpen
  \bibfield  {author} {\bibinfo {author} {\bibfnamefont {M.}~\bibnamefont
  {Piccin}}, \bibinfo {author} {\bibfnamefont {G.}~\bibnamefont {Bais}},
  \bibinfo {author} {\bibfnamefont {V.}~\bibnamefont {Grillo}}, \bibinfo
  {author} {\bibfnamefont {F.}~\bibnamefont {Jabeen}}, \bibinfo {author}
  {\bibfnamefont {S.~D.}\ \bibnamefont {Franceschi}}, \bibinfo {author}
  {\bibfnamefont {E.}~\bibnamefont {Carlino}}, \bibinfo {author} {\bibfnamefont
  {M.}~\bibnamefont {Lazzarino}}, \bibinfo {author} {\bibfnamefont
  {F.}~\bibnamefont {Romanato}}, \bibinfo {author} {\bibfnamefont
  {L.}~\bibnamefont {Businaro}}, \bibinfo {author} {\bibfnamefont {S.~R.}\
  \bibnamefont {F.}}, \bibinfo {author} {\bibnamefont {Martelli}}, \ and\
  \bibinfo {author} {\bibfnamefont {A.}~\bibnamefont {Franciosi}},\ }\href@noop
  {} {\bibfield  {journal} {\bibinfo  {journal} {Physica E}\ }\textbf {\bibinfo
  {volume} {37}},\ \bibinfo {pages} {134} (\bibinfo {year} {2007})}\BibitemShut
  {NoStop}%
\bibitem [{\citenamefont {Tambe}\ \emph {et~al.}(2010)\citenamefont {Tambe},
  \citenamefont {Ren},\ and\ \citenamefont {Gradeak}}]{Tambe}%
  \BibitemOpen
  \bibfield  {author} {\bibinfo {author} {\bibfnamefont {M.}~\bibnamefont
  {Tambe}}, \bibinfo {author} {\bibfnamefont {S.}~\bibnamefont {Ren}}, \ and\
  \bibinfo {author} {\bibfnamefont {S.}~\bibnamefont {Gradeak}},\ }\href@noop
  {} {\bibfield  {journal} {\bibinfo  {journal} {Nano Lett.}\ }\textbf
  {\bibinfo {volume} {10}},\ \bibinfo {pages} {4584} (\bibinfo {year}
  {2010})}\BibitemShut {NoStop}%
\bibitem [{\citenamefont {Thelander}\ \emph {et~al.}(2010)\citenamefont
  {Thelander}, \citenamefont {Dick}, \citenamefont {Borgstr{\"o}m},
  \citenamefont {Fr{\"o}berg}, \citenamefont {Caroff}, \citenamefont
  {Nilsson},\ and\ \citenamefont {Samuelson}}]{ThelanderNanotech}%
  \BibitemOpen
  \bibfield  {author} {\bibinfo {author} {\bibfnamefont {C.}~\bibnamefont
  {Thelander}}, \bibinfo {author} {\bibfnamefont {K.~A.}\ \bibnamefont {Dick}},
  \bibinfo {author} {\bibfnamefont {M.}~\bibnamefont {Borgstr{\"o}m}}, \bibinfo
  {author} {\bibfnamefont {L.}~\bibnamefont {Fr{\"o}berg}}, \bibinfo {author}
  {\bibfnamefont {P.}~\bibnamefont {Caroff}}, \bibinfo {author} {\bibfnamefont
  {H.}~\bibnamefont {Nilsson}}, \ and\ \bibinfo {author} {\bibfnamefont
  {L.}~\bibnamefont {Samuelson}},\ }\href@noop {} {\bibfield  {journal}
  {\bibinfo  {journal} {Nanotechnology}\ }\textbf {\bibinfo {volume} {21}},\
  \bibinfo {pages} {205703} (\bibinfo {year} {2010})}\BibitemShut {NoStop}%
\bibitem [{\citenamefont {Ford}\ \emph {et~al.}(2010)\citenamefont {Ford},
  \citenamefont {Chuang}, \citenamefont {Ho}, \citenamefont {Chueh},
  \citenamefont {Fan},\ and\ \citenamefont {Javey}}]{Ford}%
  \BibitemOpen
  \bibfield  {author} {\bibinfo {author} {\bibfnamefont {A.}~\bibnamefont
  {Ford}}, \bibinfo {author} {\bibfnamefont {S.}~\bibnamefont {Chuang}},
  \bibinfo {author} {\bibfnamefont {J.}~\bibnamefont {Ho}}, \bibinfo {author}
  {\bibfnamefont {Y.-L.}\ \bibnamefont {Chueh}}, \bibinfo {author}
  {\bibfnamefont {Z.}~\bibnamefont {Fan}}, \ and\ \bibinfo {author}
  {\bibfnamefont {A.}~\bibnamefont {Javey}},\ }\href@noop {} {\bibfield
  {journal} {\bibinfo  {journal} {Nano Lett.}\ }\textbf {\bibinfo {volume}
  {10}},\ \bibinfo {pages} {509} (\bibinfo {year} {2010})}\BibitemShut
  {NoStop}%
\bibitem [{\citenamefont {Leao}\ \emph {et~al.}(2008)\citenamefont {Leao},
  \citenamefont {Fazzio},\ and\ \citenamefont {da~Silva}}]{Leao}%
  \BibitemOpen
  \bibfield  {author} {\bibinfo {author} {\bibfnamefont {C.}~\bibnamefont
  {Leao}}, \bibinfo {author} {\bibfnamefont {A.}~\bibnamefont {Fazzio}}, \ and\
  \bibinfo {author} {\bibfnamefont {A.}~\bibnamefont {da~Silva}},\ }\href@noop
  {} {\bibfield  {journal} {\bibinfo  {journal} {Nano Lett.}\ }\textbf
  {\bibinfo {volume} {8}},\ \bibinfo {pages} {1866} (\bibinfo {year}
  {2008})}\BibitemShut {NoStop}%
\bibitem [{\citenamefont {Rurali}\ and\ \citenamefont
  {Cartoix\`{a}}(2009)}]{Rurali}%
  \BibitemOpen
  \bibfield  {author} {\bibinfo {author} {\bibfnamefont {R.}~\bibnamefont
  {Rurali}}\ and\ \bibinfo {author} {\bibfnamefont {X.}~\bibnamefont
  {Cartoix\`{a}}},\ }\href@noop {} {\bibfield  {journal} {\bibinfo  {journal}
  {Nano Lett.}\ }\textbf {\bibinfo {volume} {9}},\ \bibinfo {pages} {975}
  (\bibinfo {year} {2009})}\BibitemShut {NoStop}%
\bibitem [{\citenamefont {Peelaers}\ \emph {et~al.}(2007)\citenamefont
  {Peelaers}, \citenamefont {B.Partoens},\ and\ \citenamefont
  {Peeters}}]{peelaers-apl}%
  \BibitemOpen
  \bibfield  {author} {\bibinfo {author} {\bibfnamefont {H.}~\bibnamefont
  {Peelaers}}, \bibinfo {author} {\bibnamefont {B.Partoens}}, \ and\ \bibinfo
  {author} {\bibfnamefont {F.}~\bibnamefont {Peeters}},\ }\href@noop {}
  {\bibfield  {journal} {\bibinfo  {journal} {Appl. Phys. Lett.}\ }\textbf
  {\bibinfo {volume} {90}},\ \bibinfo {pages} {263103} (\bibinfo {year}
  {2007})}\BibitemShut {NoStop}%
\bibitem [{\citenamefont {Ghaderi}\ \emph {et~al.}(2010)\citenamefont
  {Ghaderi}, \citenamefont {Peressi}, \citenamefont {Binggeli},\ and\
  \citenamefont {Akbarzadeh}}]{Ghaderi}%
  \BibitemOpen
  \bibfield  {author} {\bibinfo {author} {\bibfnamefont {N.}~\bibnamefont
  {Ghaderi}}, \bibinfo {author} {\bibfnamefont {M.}~\bibnamefont {Peressi}},
  \bibinfo {author} {\bibfnamefont {N.}~\bibnamefont {Binggeli}}, \ and\
  \bibinfo {author} {\bibfnamefont {H.}~\bibnamefont {Akbarzadeh}},\
  }\href@noop {} {\bibfield  {journal} {\bibinfo  {journal} {Phys. Rev. B}\
  }\textbf {\bibinfo {volume} {81}},\ \bibinfo {pages} {155311} (\bibinfo
  {year} {2010})}\BibitemShut {NoStop}%
\bibitem [{\citenamefont {dos Santos}\ \emph {et~al.}(2011)\citenamefont {dos
  Santos}, \citenamefont {Schmidt},\ and\ \citenamefont {Piquini}}]{dosSantos}%
  \BibitemOpen
  \bibfield  {author} {\bibinfo {author} {\bibfnamefont {C.~L.}\ \bibnamefont
  {dos Santos}}, \bibinfo {author} {\bibfnamefont {T.~M.}\ \bibnamefont
  {Schmidt}}, \ and\ \bibinfo {author} {\bibfnamefont {P.}~\bibnamefont
  {Piquini}},\ }\href@noop {} {\bibfield  {journal} {\bibinfo  {journal}
  {Nanotechnology}\ }\textbf {\bibinfo {volume} {22}},\ \bibinfo {pages}
  {265203} (\bibinfo {year} {2011})}\BibitemShut {NoStop}%
\bibitem [{\citenamefont {Shu}\ \emph {et~al.}(2011)\citenamefont {Shu},
  \citenamefont {Chen}, \citenamefont {Ding}, \citenamefont {Dong},\ and\
  \citenamefont {Lei}}]{Shu11}%
  \BibitemOpen
  \bibfield  {author} {\bibinfo {author} {\bibfnamefont {H.}~\bibnamefont
  {Shu}}, \bibinfo {author} {\bibfnamefont {X.}~\bibnamefont {Chen}}, \bibinfo
  {author} {\bibfnamefont {Z.}~\bibnamefont {Ding}}, \bibinfo {author}
  {\bibfnamefont {R.}~\bibnamefont {Dong}}, \ and\ \bibinfo {author}
  {\bibfnamefont {W.}~\bibnamefont {Lei}},\ }\href@noop {} {\bibfield
  {journal} {\bibinfo  {journal} {J. Phys. Chem. C}\ }\textbf {\bibinfo
  {volume} {115}},\ \bibinfo {pages} {14449} (\bibinfo {year}
  {2011})}\BibitemShut {NoStop}%
\bibitem [{\citenamefont {Shu}\ \emph {et~al.}(2012)\citenamefont {Shu},
  \citenamefont {Cao}, \citenamefont {Liang}, \citenamefont {Jin},
  \citenamefont {Chen},\ and\ \citenamefont {Lei}}]{Shu12}%
  \BibitemOpen
  \bibfield  {author} {\bibinfo {author} {\bibfnamefont {H.}~\bibnamefont
  {Shu}}, \bibinfo {author} {\bibfnamefont {D.}~\bibnamefont {Cao}}, \bibinfo
  {author} {\bibfnamefont {P.}~\bibnamefont {Liang}}, \bibinfo {author}
  {\bibfnamefont {S.}~\bibnamefont {Jin}}, \bibinfo {author} {\bibfnamefont
  {X.}~\bibnamefont {Chen}}, \ and\ \bibinfo {author} {\bibfnamefont
  {W.}~\bibnamefont {Lei}},\ }\href@noop {} {\bibfield  {journal} {\bibinfo
  {journal} {J. Phys. Chem. C}\ }\textbf {\bibinfo {volume} {116}},\ \bibinfo
  {pages} {17928} (\bibinfo {year} {2012})}\BibitemShut {NoStop}%
\bibitem [{\citenamefont {Hasegawa}\ and\ \citenamefont
  {Akazawa}(2008)}]{Hasegawa}%
  \BibitemOpen
  \bibfield  {author} {\bibinfo {author} {\bibfnamefont {H.}~\bibnamefont
  {Hasegawa}}\ and\ \bibinfo {author} {\bibfnamefont {M.}~\bibnamefont
  {Akazawa}},\ }\href@noop {} {\bibfield  {journal} {\bibinfo  {journal} {Appl.
  Surf. Sci.}\ }\textbf {\bibinfo {volume} {255}},\ \bibinfo {pages} {628}
  (\bibinfo {year} {2008})}\BibitemShut {NoStop}%
\bibitem [{\citenamefont {Czaban}\ \emph {et~al.}(2009)\citenamefont {Czaban},
  \citenamefont {Thompson},\ and\ \citenamefont {LaPierre}}]{Czaban}%
  \BibitemOpen
  \bibfield  {author} {\bibinfo {author} {\bibfnamefont {J.}~\bibnamefont
  {Czaban}}, \bibinfo {author} {\bibfnamefont {D.}~\bibnamefont {Thompson}}, \
  and\ \bibinfo {author} {\bibfnamefont {R.}~\bibnamefont {LaPierre}},\
  }\href@noop {} {\bibfield  {journal} {\bibinfo  {journal} {Nano Lett.}\
  }\textbf {\bibinfo {volume} {9}},\ \bibinfo {pages} {148} (\bibinfo {year}
  {2009})}\BibitemShut {NoStop}%
\bibitem [{\citenamefont {Colombo}\ \emph {et~al.}(2009)\citenamefont
  {Colombo}, \citenamefont {Hei{\ss}}, \citenamefont {Gr{\"{a}}tzel},\ and\
  \citenamefont {i~Morral}}]{Colombo}%
  \BibitemOpen
  \bibfield  {author} {\bibinfo {author} {\bibfnamefont {C.}~\bibnamefont
  {Colombo}}, \bibinfo {author} {\bibfnamefont {M.}~\bibnamefont {Hei{\ss}}},
  \bibinfo {author} {\bibfnamefont {M.}~\bibnamefont {Gr{\"{a}}tzel}}, \ and\
  \bibinfo {author} {\bibfnamefont {A.~F.}\ \bibnamefont {i~Morral}},\
  }\href@noop {} {\bibfield  {journal} {\bibinfo  {journal} {Appl. Phys.
  Lett.}\ }\textbf {\bibinfo {volume} {94}},\ \bibinfo {pages} {173108}
  (\bibinfo {year} {2009})}\BibitemShut {NoStop}%
\bibitem [{\citenamefont {Dufouleur}\ \emph {et~al.}(2010)\citenamefont
  {Dufouleur}, \citenamefont {Colombo}, \citenamefont {Garma}, \citenamefont
  {Ketterer}, \citenamefont {Uccelli}, \citenamefont {Nicotra},\ and\
  \citenamefont {i~Morral}}]{Dufouleur}%
  \BibitemOpen
  \bibfield  {author} {\bibinfo {author} {\bibfnamefont {J.}~\bibnamefont
  {Dufouleur}}, \bibinfo {author} {\bibfnamefont {C.}~\bibnamefont {Colombo}},
  \bibinfo {author} {\bibfnamefont {T.}~\bibnamefont {Garma}}, \bibinfo
  {author} {\bibfnamefont {B.}~\bibnamefont {Ketterer}}, \bibinfo {author}
  {\bibfnamefont {E.}~\bibnamefont {Uccelli}}, \bibinfo {author} {\bibfnamefont
  {M.}~\bibnamefont {Nicotra}}, \ and\ \bibinfo {author} {\bibfnamefont
  {A.~F.}\ \bibnamefont {i~Morral}},\ }\href@noop {} {\bibfield  {journal}
  {\bibinfo  {journal} {Nano Lett.}\ }\textbf {\bibinfo {volume} {10}},\
  \bibinfo {pages} {1734} (\bibinfo {year} {2010})}\BibitemShut {NoStop}%
\bibitem [{\citenamefont {Hilse}\ \emph {et~al.}(2010)\citenamefont {Hilse},
  \citenamefont {Ramsteiner}, \citenamefont {Breuer}, \citenamefont
  {Geelhaar},\ and\ \citenamefont {Riechert}}]{hilse-si}%
  \BibitemOpen
  \bibfield  {author} {\bibinfo {author} {\bibfnamefont {M.}~\bibnamefont
  {Hilse}}, \bibinfo {author} {\bibfnamefont {M.}~\bibnamefont {Ramsteiner}},
  \bibinfo {author} {\bibfnamefont {S.}~\bibnamefont {Breuer}}, \bibinfo
  {author} {\bibfnamefont {L.}~\bibnamefont {Geelhaar}}, \ and\ \bibinfo
  {author} {\bibfnamefont {H.}~\bibnamefont {Riechert}},\ }\href@noop {}
  {\bibfield  {journal} {\bibinfo  {journal} {Appl. Phys. Lett.}\ }\textbf
  {\bibinfo {volume} {96}},\ \bibinfo {pages} {193104} (\bibinfo {year}
  {2010})}\BibitemShut {NoStop}%
\bibitem [{\citenamefont {Gutsche}\ \emph {et~al.}(2011)\citenamefont
  {Gutsche}, \citenamefont {Lysov}, \citenamefont {Regolin}, \citenamefont
  {Blekker}, \citenamefont {Prost},\ and\ \citenamefont {Tegude}}]{Gutsche}%
  \BibitemOpen
  \bibfield  {author} {\bibinfo {author} {\bibfnamefont {C.}~\bibnamefont
  {Gutsche}}, \bibinfo {author} {\bibfnamefont {A.}~\bibnamefont {Lysov}},
  \bibinfo {author} {\bibfnamefont {I.}~\bibnamefont {Regolin}}, \bibinfo
  {author} {\bibfnamefont {K.}~\bibnamefont {Blekker}}, \bibinfo {author}
  {\bibfnamefont {W.}~\bibnamefont {Prost}}, \ and\ \bibinfo {author}
  {\bibfnamefont {F.-J.}\ \bibnamefont {Tegude}},\ }\href@noop {} {\bibfield
  {journal} {\bibinfo  {journal} {Nanoscale Res. Lett.}\ }\textbf {\bibinfo
  {volume} {6}},\ \bibinfo {pages} {65} (\bibinfo {year} {2011})}\BibitemShut
  {NoStop}%
\bibitem [{\citenamefont {Erwin}\ \emph {et~al.}(2005)\citenamefont {Erwin},
  \citenamefont {Zu}, \citenamefont {Haftel}, \citenamefont {Efros},
  \citenamefont {Kennedy},\ and\ \citenamefont {Norris}}]{Erwin}%
  \BibitemOpen
  \bibfield  {author} {\bibinfo {author} {\bibfnamefont {S.}~\bibnamefont
  {Erwin}}, \bibinfo {author} {\bibfnamefont {L.}~\bibnamefont {Zu}}, \bibinfo
  {author} {\bibfnamefont {M.~I.}\ \bibnamefont {Haftel}}, \bibinfo {author}
  {\bibfnamefont {A.}~\bibnamefont {Efros}}, \bibinfo {author} {\bibfnamefont
  {T.}~\bibnamefont {Kennedy}}, \ and\ \bibinfo {author} {\bibfnamefont
  {D.}~\bibnamefont {Norris}},\ }\href@noop {} {\bibfield  {journal} {\bibinfo
  {journal} {Nature (London)}\ }\textbf {\bibinfo {volume} {436}},\ \bibinfo
  {pages} {91} (\bibinfo {year} {2005})}\BibitemShut {NoStop}%
\bibitem [{\citenamefont {Chen}\ \emph {et~al.}(2012)\citenamefont {Chen},
  \citenamefont {Dubrovskii}, \citenamefont {Liu}, \citenamefont {Xu},
  \citenamefont {Larde}, \citenamefont {Nys}, \citenamefont {Grandidier},
  \citenamefont {Stievenard}, \citenamefont {Patriarche},\ and\ \citenamefont
  {Pareige}}]{Chen}%
  \BibitemOpen
  \bibfield  {author} {\bibinfo {author} {\bibfnamefont {W.}~\bibnamefont
  {Chen}}, \bibinfo {author} {\bibfnamefont {V.~G.}\ \bibnamefont
  {Dubrovskii}}, \bibinfo {author} {\bibfnamefont {X.}~\bibnamefont {Liu}},
  \bibinfo {author} {\bibfnamefont {T.}~\bibnamefont {Xu}}, \bibinfo {author}
  {\bibfnamefont {R.}~\bibnamefont {Larde}}, \bibinfo {author} {\bibfnamefont
  {J.~P.}\ \bibnamefont {Nys}}, \bibinfo {author} {\bibfnamefont
  {B.}~\bibnamefont {Grandidier}}, \bibinfo {author} {\bibfnamefont
  {D.}~\bibnamefont {Stievenard}}, \bibinfo {author} {\bibfnamefont
  {G.}~\bibnamefont {Patriarche}}, \ and\ \bibinfo {author} {\bibfnamefont
  {P.}~\bibnamefont {Pareige}},\ }\href {\doibase 10.1063/1.4714364} {\bibfield
   {journal} {\bibinfo  {journal} {J. Appl. Phys.}\ }\textbf {\bibinfo {volume}
  {111}},\ \bibinfo {pages} {094909} (\bibinfo {year} {2012})}\BibitemShut
  {NoStop}%
\bibitem [{\citenamefont {Galicka}\ \emph {et~al.}(2011)\citenamefont
  {Galicka}, \citenamefont {Buczko},\ and\ \citenamefont {Kacman}}]{galicka2}%
  \BibitemOpen
  \bibfield  {author} {\bibinfo {author} {\bibfnamefont {M.}~\bibnamefont
  {Galicka}}, \bibinfo {author} {\bibfnamefont {R.}~\bibnamefont {Buczko}}, \
  and\ \bibinfo {author} {\bibfnamefont {P.}~\bibnamefont {Kacman}},\
  }\href@noop {} {\bibfield  {journal} {\bibinfo  {journal} {Nano Lett.}\
  }\textbf {\bibinfo {volume} {11}},\ \bibinfo {pages} {3319} (\bibinfo {year}
  {2011})}\BibitemShut {NoStop}%
\bibitem [{\citenamefont {Huang}\ and\ \citenamefont {Kuech}(1994)}]{Huang}%
  \BibitemOpen
  \bibfield  {author} {\bibinfo {author} {\bibfnamefont {J.~W.}\ \bibnamefont
  {Huang}}\ and\ \bibinfo {author} {\bibfnamefont {T.~F.}\ \bibnamefont
  {Kuech}},\ }\href@noop {} {\bibfield  {journal} {\bibinfo  {journal} {Appl.
  Phys. Lett.}\ }\textbf {\bibinfo {volume} {65}},\ \bibinfo {pages} {604}
  (\bibinfo {year} {1994})}\BibitemShut {NoStop}%
\bibitem [{\citenamefont {Salem}\ \emph {et~al.}(2006)\citenamefont {Salem},
  \citenamefont {Morris}, \citenamefont {Salissou}, \citenamefont {Aimez},
  \citenamefont {Charlebois}, \citenamefont {Chicoine},\ and\ \citenamefont
  {Schiettekatte}}]{Salem}%
  \BibitemOpen
  \bibfield  {author} {\bibinfo {author} {\bibfnamefont {B.}~\bibnamefont
  {Salem}}, \bibinfo {author} {\bibfnamefont {D.}~\bibnamefont {Morris}},
  \bibinfo {author} {\bibfnamefont {Y.}~\bibnamefont {Salissou}}, \bibinfo
  {author} {\bibfnamefont {V.}~\bibnamefont {Aimez}}, \bibinfo {author}
  {\bibfnamefont {S.}~\bibnamefont {Charlebois}}, \bibinfo {author}
  {\bibfnamefont {M.}~\bibnamefont {Chicoine}}, \ and\ \bibinfo {author}
  {\bibfnamefont {F.}~\bibnamefont {Schiettekatte}},\ }\href@noop {} {\bibfield
   {journal} {\bibinfo  {journal} {J. Vac. Sci. Technol. A}\ }\textbf {\bibinfo
  {volume} {24}},\ \bibinfo {pages} {774} (\bibinfo {year} {2006})}\BibitemShut
  {NoStop}%
\bibitem [{\citenamefont {Zhi}\ \emph {et~al.}(2005)\citenamefont {Zhi},
  \citenamefont {Bai},\ and\ \citenamefont {Wang}}]{Zhi}%
  \BibitemOpen
  \bibfield  {author} {\bibinfo {author} {\bibfnamefont {C.~Y.}\ \bibnamefont
  {Zhi}}, \bibinfo {author} {\bibfnamefont {X.~D.}\ \bibnamefont {Bai}}, \ and\
  \bibinfo {author} {\bibfnamefont {E.~G.}\ \bibnamefont {Wang}},\ }\href@noop
  {} {\bibfield  {journal} {\bibinfo  {journal} {Appl. Phys. Lett.}\ }\textbf
  {\bibinfo {volume} {86}},\ \bibinfo {pages} {213108} (\bibinfo {year}
  {2005})}\BibitemShut {NoStop}%
\bibitem [{\citenamefont {Allen}\ \emph {et~al.}(2008)\citenamefont {Allen},
  \citenamefont {Hemesath}, \citenamefont {Perea}, \citenamefont {Lensch-Falk},
  \citenamefont {Li}, \citenamefont {Yin}, \citenamefont {Gass}, \citenamefont
  {Wang}, \citenamefont {Bleloch}, \citenamefont {Palmer},\ and\ \citenamefont
  {Lauhon}}]{Allen}%
  \BibitemOpen
  \bibfield  {author} {\bibinfo {author} {\bibfnamefont {J.~E.}\ \bibnamefont
  {Allen}}, \bibinfo {author} {\bibfnamefont {E.~R.}\ \bibnamefont {Hemesath}},
  \bibinfo {author} {\bibfnamefont {D.~E.}\ \bibnamefont {Perea}}, \bibinfo
  {author} {\bibfnamefont {J.~L.}\ \bibnamefont {Lensch-Falk}}, \bibinfo
  {author} {\bibfnamefont {Z.~Y.}\ \bibnamefont {Li}}, \bibinfo {author}
  {\bibfnamefont {F.}~\bibnamefont {Yin}}, \bibinfo {author} {\bibfnamefont
  {M.~H.}\ \bibnamefont {Gass}}, \bibinfo {author} {\bibfnamefont
  {P.}~\bibnamefont {Wang}}, \bibinfo {author} {\bibfnamefont {A.~L.}\
  \bibnamefont {Bleloch}}, \bibinfo {author} {\bibfnamefont {R.~E.}\
  \bibnamefont {Palmer}}, \ and\ \bibinfo {author} {\bibfnamefont {L.~J.}\
  \bibnamefont {Lauhon}},\ }\href@noop {} {\bibfield  {journal} {\bibinfo
  {journal} {Nat. Nanotechnol.}\ }\textbf {\bibinfo {volume} {3}},\ \bibinfo
  {pages} {168} (\bibinfo {year} {2008})}\BibitemShut {NoStop}%
\bibitem [{\citenamefont {Perea}\ \emph {et~al.}(2006)\citenamefont {Perea},
  \citenamefont {Allen}, \citenamefont {May}, \citenamefont {Wessels},
  \citenamefont {Seidman},\ and\ \citenamefont {Lauhon}}]{Perea}%
  \BibitemOpen
  \bibfield  {author} {\bibinfo {author} {\bibfnamefont {D.~E.}\ \bibnamefont
  {Perea}}, \bibinfo {author} {\bibfnamefont {J.~E.}\ \bibnamefont {Allen}},
  \bibinfo {author} {\bibfnamefont {S.~J.}\ \bibnamefont {May}}, \bibinfo
  {author} {\bibfnamefont {B.~W.}\ \bibnamefont {Wessels}}, \bibinfo {author}
  {\bibfnamefont {D.~N.}\ \bibnamefont {Seidman}}, \ and\ \bibinfo {author}
  {\bibfnamefont {L.~J.}\ \bibnamefont {Lauhon}},\ }\href@noop {} {\bibfield
  {journal} {\bibinfo  {journal} {Nano Lett.}\ }\textbf {\bibinfo {volume}
  {6}},\ \bibinfo {pages} {181} (\bibinfo {year} {2006})}\BibitemShut {NoStop}%
\bibitem [{\citenamefont {Bar-Sadan}\ \emph {et~al.}(2012)\citenamefont
  {Bar-Sadan}, \citenamefont {Barthel}, \citenamefont {Shtrikman},\ and\
  \citenamefont {Popovitz-Biro}}]{Hadasgold}%
  \BibitemOpen
  \bibfield  {author} {\bibinfo {author} {\bibfnamefont {M.}~\bibnamefont
  {Bar-Sadan}}, \bibinfo {author} {\bibfnamefont {J.}~\bibnamefont {Barthel}},
  \bibinfo {author} {\bibfnamefont {H.}~\bibnamefont {Shtrikman}}, \ and\
  \bibinfo {author} {\bibfnamefont {R.}~\bibnamefont {Popovitz-Biro}},\
  }\href@noop {} {\bibfield  {journal} {\bibinfo  {journal} {Nano Lett.}\
  }\textbf {\bibinfo {volume} {12}},\ \bibinfo {pages} {2352} (\bibinfo {year}
  {2012})}\BibitemShut {NoStop}%
\bibitem [{\citenamefont {Mandl}\ \emph {et~al.}(2006)\citenamefont {Mandl},
  \citenamefont {Stangl}, \citenamefont {M{\aa}rtensson}, \citenamefont
  {Mikkelsen}, \citenamefont {Eriksson}, \citenamefont {Karlsson},
  \citenamefont {abd L.~Samuelson},\ and\ \citenamefont {Seifert}}]{Mandl2006}%
  \BibitemOpen
  \bibfield  {author} {\bibinfo {author} {\bibfnamefont {B.}~\bibnamefont
  {Mandl}}, \bibinfo {author} {\bibfnamefont {J.}~\bibnamefont {Stangl}},
  \bibinfo {author} {\bibfnamefont {T.}~\bibnamefont {M{\aa}rtensson}},
  \bibinfo {author} {\bibfnamefont {A.}~\bibnamefont {Mikkelsen}}, \bibinfo
  {author} {\bibfnamefont {J.}~\bibnamefont {Eriksson}}, \bibinfo {author}
  {\bibfnamefont {L.~S.}\ \bibnamefont {Karlsson}}, \bibinfo {author}
  {\bibfnamefont {G.~U. N.~B.}\ \bibnamefont {abd L.~Samuelson}}, \ and\
  \bibinfo {author} {\bibfnamefont {W.}~\bibnamefont {Seifert}},\ }\href@noop
  {} {\enquote {\bibinfo {title} {Au-free epitaxial growth of inas
  nanowires},}\ } (\bibinfo {year} {2006})\BibitemShut {NoStop}%
\bibitem [{\citenamefont {Jabeen}\ \emph {et~al.}(2008)\citenamefont {Jabeen},
  \citenamefont {Grillo}, \citenamefont {Rubibi},\ and\ \citenamefont
  {Martelli}}]{Jabeen}%
  \BibitemOpen
  \bibfield  {author} {\bibinfo {author} {\bibfnamefont {F.}~\bibnamefont
  {Jabeen}}, \bibinfo {author} {\bibfnamefont {V.}~\bibnamefont {Grillo}},
  \bibinfo {author} {\bibfnamefont {S.}~\bibnamefont {Rubibi}}, \ and\ \bibinfo
  {author} {\bibfnamefont {F.}~\bibnamefont {Martelli}},\ }\href@noop {}
  {\bibfield  {journal} {\bibinfo  {journal} {Nanotechnology}\ }\textbf
  {\bibinfo {volume} {19}},\ \bibinfo {pages} {275711} (\bibinfo {year}
  {2008})}\BibitemShut {NoStop}%
\bibitem [{\citenamefont {Hei{\ss}}\ \emph {et~al.}(2009)\citenamefont
  {Hei{\ss}}, \citenamefont {Gustafsson}, \citenamefont {Conesa-Boj},
  \citenamefont {Peir\'{o}}, \citenamefont {Morante}, \citenamefont
  {Abstreiter}, \citenamefont {Arbiol}, \citenamefont {Samuelson},\ and\
  \citenamefont {i~Morral}}]{AFMorral}%
  \BibitemOpen
  \bibfield  {author} {\bibinfo {author} {\bibfnamefont {M.}~\bibnamefont
  {Hei{\ss}}}, \bibinfo {author} {\bibfnamefont {A.}~\bibnamefont
  {Gustafsson}}, \bibinfo {author} {\bibfnamefont {S.}~\bibnamefont
  {Conesa-Boj}}, \bibinfo {author} {\bibfnamefont {F.}~\bibnamefont
  {Peir\'{o}}}, \bibinfo {author} {\bibfnamefont {J.~R.}\ \bibnamefont
  {Morante}}, \bibinfo {author} {\bibfnamefont {G.}~\bibnamefont {Abstreiter}},
  \bibinfo {author} {\bibfnamefont {J.}~\bibnamefont {Arbiol}}, \bibinfo
  {author} {\bibfnamefont {L.}~\bibnamefont {Samuelson}}, \ and\ \bibinfo
  {author} {\bibfnamefont {A.~F.}\ \bibnamefont {i~Morral}},\ }\href@noop {} {}
  (\bibinfo {year} {2009})\BibitemShut {NoStop}%
\bibitem [{\citenamefont {Krogstrup}\ \emph {et~al.}(2010)\citenamefont
  {Krogstrup}, \citenamefont {Popovitz-Biro}, \citenamefont {Johnson},
  \citenamefont {Madsen}, \citenamefont {Nygard},\ and\ \citenamefont
  {Shtrikman}}]{Peter}%
  \BibitemOpen
  \bibfield  {author} {\bibinfo {author} {\bibfnamefont {P.}~\bibnamefont
  {Krogstrup}}, \bibinfo {author} {\bibfnamefont {R.}~\bibnamefont
  {Popovitz-Biro}}, \bibinfo {author} {\bibfnamefont {E.}~\bibnamefont
  {Johnson}}, \bibinfo {author} {\bibfnamefont {M.~H.}\ \bibnamefont {Madsen}},
  \bibinfo {author} {\bibfnamefont {J.}~\bibnamefont {Nygard}}, \ and\ \bibinfo
  {author} {\bibfnamefont {H.}~\bibnamefont {Shtrikman}},\ }\href@noop {}
  {\bibfield  {journal} {\bibinfo  {journal} {Nano Lett.}\ }\textbf {\bibinfo
  {volume} {10}},\ \bibinfo {pages} {4475} (\bibinfo {year}
  {2010})}\BibitemShut {NoStop}%
\bibitem [{\citenamefont {Northrup}\ and\ \citenamefont
  {Zhang}(1993)}]{Northrup}%
  \BibitemOpen
  \bibfield  {author} {\bibinfo {author} {\bibfnamefont {J.~E.}\ \bibnamefont
  {Northrup}}\ and\ \bibinfo {author} {\bibfnamefont {S.~B.}\ \bibnamefont
  {Zhang}},\ }\href@noop {} {\bibfield  {journal} {\bibinfo  {journal} {Phys.
  Rev. B}\ }\textbf {\bibinfo {volume} {47}},\ \bibinfo {pages} {6791}
  (\bibinfo {year} {1993})}\BibitemShut {NoStop}%
\bibitem [{\citenamefont {Galicka}\ \emph {et~al.}(2008)\citenamefont
  {Galicka}, \citenamefont {Bukala}, \citenamefont {Buczko},\ and\
  \citenamefont {Kacman}}]{JPCM}%
  \BibitemOpen
  \bibfield  {author} {\bibinfo {author} {\bibfnamefont {M.}~\bibnamefont
  {Galicka}}, \bibinfo {author} {\bibfnamefont {M.}~\bibnamefont {Bukala}},
  \bibinfo {author} {\bibfnamefont {R.}~\bibnamefont {Buczko}}, \ and\ \bibinfo
  {author} {\bibfnamefont {P.}~\bibnamefont {Kacman}},\ }\href@noop {}
  {\bibfield  {journal} {\bibinfo  {journal} {J. Phys.: Condens. Matter}\
  }\textbf {\bibinfo {volume} {20}},\ \bibinfo {pages} {454226} (\bibinfo
  {year} {2008})}\BibitemShut {NoStop}%
\bibitem [{\citenamefont {Glas}\ \emph {et~al.}(2007)\citenamefont {Glas},
  \citenamefont {Harmand},\ and\ \citenamefont {Patriarche}}]{Glas}%
  \BibitemOpen
  \bibfield  {author} {\bibinfo {author} {\bibfnamefont {F.}~\bibnamefont
  {Glas}}, \bibinfo {author} {\bibfnamefont {J.-C.}\ \bibnamefont {Harmand}}, \
  and\ \bibinfo {author} {\bibfnamefont {G.}~\bibnamefont {Patriarche}},\
  }\href@noop {} {\bibfield  {journal} {\bibinfo  {journal} {Phys. Rev. Lett.}\
  }\textbf {\bibinfo {volume} {99}},\ \bibinfo {pages} {146101} (\bibinfo
  {year} {2007})}\BibitemShut {NoStop}%
\bibitem [{\citenamefont {Dubrovskii}\ \emph {et~al.}(2008)\citenamefont
  {Dubrovskii}, \citenamefont {Sibirev}, \citenamefont {Harmand},\ and\
  \citenamefont {Glas}}]{dubrowski}%
  \BibitemOpen
  \bibfield  {author} {\bibinfo {author} {\bibfnamefont {V.~G.}\ \bibnamefont
  {Dubrovskii}}, \bibinfo {author} {\bibfnamefont {N.}~\bibnamefont {Sibirev}},
  \bibinfo {author} {\bibfnamefont {J.-C.}\ \bibnamefont {Harmand}}, \ and\
  \bibinfo {author} {\bibfnamefont {F.}~\bibnamefont {Glas}},\ }\href@noop {}
  {\bibfield  {journal} {\bibinfo  {journal} {Phys. Rev. B}\ }\textbf {\bibinfo
  {volume} {78}},\ \bibinfo {pages} {235301} (\bibinfo {year}
  {2008})}\BibitemShut {NoStop}%
\bibitem [{\citenamefont {Cahangirov}\ and\ \citenamefont
  {Ciraci}(2009)}]{Ciraci}%
  \BibitemOpen
  \bibfield  {author} {\bibinfo {author} {\bibfnamefont {S.}~\bibnamefont
  {Cahangirov}}\ and\ \bibinfo {author} {\bibfnamefont {S.}~\bibnamefont
  {Ciraci}},\ }\href@noop {} {\bibfield  {journal} {\bibinfo  {journal} {Phys.
  Rev. B}\ }\textbf {\bibinfo {volume} {79}},\ \bibinfo {pages} {165118}
  (\bibinfo {year} {2009})}\BibitemShut {NoStop}%
\bibitem [{\citenamefont {Shtrikman}\ \emph {et~al.}(2009)\citenamefont
  {Shtrikman}, \citenamefont {Popovitz-Biro}, \citenamefont {Kretinin},
  \citenamefont {Houben}, \citenamefont {Heiblum}, \citenamefont {Bukala},
  \citenamefont {Galicka}, \citenamefont {Buczko},\ and\ \citenamefont
  {Kacman}}]{Shtrikman}%
  \BibitemOpen
  \bibfield  {author} {\bibinfo {author} {\bibfnamefont {H.}~\bibnamefont
  {Shtrikman}}, \bibinfo {author} {\bibfnamefont {R.}~\bibnamefont
  {Popovitz-Biro}}, \bibinfo {author} {\bibfnamefont {A.}~\bibnamefont
  {Kretinin}}, \bibinfo {author} {\bibfnamefont {L.}~\bibnamefont {Houben}},
  \bibinfo {author} {\bibfnamefont {M.}~\bibnamefont {Heiblum}}, \bibinfo
  {author} {\bibfnamefont {M.}~\bibnamefont {Bukala}}, \bibinfo {author}
  {\bibfnamefont {M.}~\bibnamefont {Galicka}}, \bibinfo {author} {\bibfnamefont
  {R.}~\bibnamefont {Buczko}}, \ and\ \bibinfo {author} {\bibfnamefont
  {P.}~\bibnamefont {Kacman}},\ }\href@noop {} {\bibfield  {journal} {\bibinfo
  {journal} {Nano Lett.}\ }\textbf {\bibinfo {volume} {9}},\ \bibinfo {pages}
  {1509} (\bibinfo {year} {2009})}\BibitemShut {NoStop}%
\bibitem [{\citenamefont {Kresse}\ and\ \citenamefont {Hafner}(1993)}]{nw4}%
  \BibitemOpen
  \bibfield  {author} {\bibinfo {author} {\bibfnamefont {G.}~\bibnamefont
  {Kresse}}\ and\ \bibinfo {author} {\bibfnamefont {J.}~\bibnamefont
  {Hafner}},\ }\href@noop {} {\bibfield  {journal} {\bibinfo  {journal} {Phys.
  Rev. B}\ }\textbf {\bibinfo {volume} {47}},\ \bibinfo {pages} {R558}
  (\bibinfo {year} {1993})}\BibitemShut {NoStop}%
\bibitem [{\citenamefont {Kresse}\ and\ \citenamefont
  {Furthm{\"{u}}ller}(1996)}]{nw5}%
  \BibitemOpen
  \bibfield  {author} {\bibinfo {author} {\bibfnamefont {G.}~\bibnamefont
  {Kresse}}\ and\ \bibinfo {author} {\bibfnamefont {J.}~\bibnamefont
  {Furthm{\"{u}}ller}},\ }\href@noop {} {\bibfield  {journal} {\bibinfo
  {journal} {Phys. Rev. B}\ }\textbf {\bibinfo {volume} {54}},\ \bibinfo
  {pages} {11169} (\bibinfo {year} {1996})}\BibitemShut {NoStop}%
\bibitem [{\citenamefont {Bl{\"{o}}chl}(1994)}]{PAW}%
  \BibitemOpen
  \bibfield  {author} {\bibinfo {author} {\bibfnamefont {P.~E.}\ \bibnamefont
  {Bl{\"{o}}chl}},\ }\href@noop {} {\bibfield  {journal} {\bibinfo  {journal}
  {Phys. Rev. B}\ }\textbf {\bibinfo {volume} {50}},\ \bibinfo {pages} {17953}
  (\bibinfo {year} {1994})}\BibitemShut {NoStop}%
\bibitem [{\citenamefont {Sadowski}\ and\ \citenamefont
  {Ramprasad}(2010)}]{TSadowski}%
  \BibitemOpen
  \bibfield  {author} {\bibinfo {author} {\bibfnamefont {T.}~\bibnamefont
  {Sadowski}}\ and\ \bibinfo {author} {\bibfnamefont {R.}~\bibnamefont
  {Ramprasad}},\ }\href@noop {} {} (\bibinfo {year} {2010})\BibitemShut
  {NoStop}%
\bibitem [{\citenamefont {Casadei}\ \emph {et~al.}(2013)\citenamefont
  {Casadei}, \citenamefont {Krogstrup}, \citenamefont {Heiss}, \citenamefont
  {R{\"o}hr}, \citenamefont {Colombo}, \citenamefont {Ruelle}, \citenamefont
  {Upadhyay}, \citenamefont {S{{\o}}rensen}, \citenamefont {Nyg{\aa}rd},\ and\
  \citenamefont {i~Morral}}]{casadei}%
  \BibitemOpen
  \bibfield  {author} {\bibinfo {author} {\bibfnamefont {A.}~\bibnamefont
  {Casadei}}, \bibinfo {author} {\bibfnamefont {P.}~\bibnamefont {Krogstrup}},
  \bibinfo {author} {\bibfnamefont {M.}~\bibnamefont {Heiss}}, \bibinfo
  {author} {\bibfnamefont {J.~A.}\ \bibnamefont {R{\"o}hr}}, \bibinfo {author}
  {\bibfnamefont {C.}~\bibnamefont {Colombo}}, \bibinfo {author} {\bibfnamefont
  {T.}~\bibnamefont {Ruelle}}, \bibinfo {author} {\bibfnamefont
  {S.}~\bibnamefont {Upadhyay}}, \bibinfo {author} {\bibfnamefont {C.~B.}\
  \bibnamefont {S{{\o}}rensen}}, \bibinfo {author} {\bibfnamefont
  {J.}~\bibnamefont {Nyg{\aa}rd}}, \ and\ \bibinfo {author} {\bibfnamefont
  {A.~F.}\ \bibnamefont {i~Morral}},\ }\href@noop {} {\bibfield  {journal}
  {\bibinfo  {journal} {Appl. Phys. Lett.}\ }\textbf {\bibinfo {volume}
  {102}},\ \bibinfo {pages} {013117} (\bibinfo {year} {2013})}\BibitemShut
  {NoStop}%
\bibitem [{\citenamefont {S{{\o}}rensen}\ \emph {et~al.}(2008)\citenamefont
  {S{{\o}}rensen}, \citenamefont {Aagesen}, \citenamefont {S{{\o}}rensen},
  \citenamefont {Lindelof}, \citenamefont {Martinez},\ and\ \citenamefont
  {Nyg{\aa}rd}}]{Sorensen}%
  \BibitemOpen
  \bibfield  {author} {\bibinfo {author} {\bibfnamefont {B.~S.}\ \bibnamefont
  {S{{\o}}rensen}}, \bibinfo {author} {\bibfnamefont {M.}~\bibnamefont
  {Aagesen}}, \bibinfo {author} {\bibfnamefont {C.~B.}\ \bibnamefont
  {S{{\o}}rensen}}, \bibinfo {author} {\bibfnamefont {P.~E.}\ \bibnamefont
  {Lindelof}}, \bibinfo {author} {\bibfnamefont {K.~L.}\ \bibnamefont
  {Martinez}}, \ and\ \bibinfo {author} {\bibfnamefont {J.}~\bibnamefont
  {Nyg{\aa}rd}},\ }\href@noop {} {\bibfield  {journal} {\bibinfo  {journal}
  {Applied Physics Letters}\ }\textbf {\bibinfo {volume} {92}},\ \bibinfo
  {pages} {012119} (\bibinfo {year} {2008})}\BibitemShut {NoStop}%
\bibitem [{\citenamefont {Astromskas}\ \emph {et~al.}(2010)\citenamefont
  {Astromskas}, \citenamefont {Storm}, \citenamefont {Karlstr{\"o}m},
  \citenamefont {Caroff}, \citenamefont {Borgstr{\"o}m},\ and\ \citenamefont
  {Wernersson}}]{Astromskas}%
  \BibitemOpen
  \bibfield  {author} {\bibinfo {author} {\bibfnamefont {G.}~\bibnamefont
  {Astromskas}}, \bibinfo {author} {\bibfnamefont {K.}~\bibnamefont {Storm}},
  \bibinfo {author} {\bibfnamefont {O.}~\bibnamefont {Karlstr{\"o}m}}, \bibinfo
  {author} {\bibfnamefont {P.}~\bibnamefont {Caroff}}, \bibinfo {author}
  {\bibfnamefont {M.}~\bibnamefont {Borgstr{\"o}m}}, \ and\ \bibinfo {author}
  {\bibfnamefont {L.-E.}\ \bibnamefont {Wernersson}},\ }\href@noop {}
  {\bibfield  {journal} {\bibinfo  {journal} {J. Appl. Phys.}\ }\textbf
  {\bibinfo {volume} {108}},\ \bibinfo {pages} {054306} (\bibinfo {year}
  {2010})}\BibitemShut {NoStop}%
\bibitem [{\citenamefont {Perea}\ \emph {et~al.}(2009)\citenamefont {Perea},
  \citenamefont {Hemesath}, \citenamefont {Schwalbach}, \citenamefont
  {Lensch-Falk}, \citenamefont {Voorhees},\ and\ \citenamefont
  {Lauhon}}]{Lauhon}%
  \BibitemOpen
  \bibfield  {author} {\bibinfo {author} {\bibfnamefont {D.~E.}\ \bibnamefont
  {Perea}}, \bibinfo {author} {\bibfnamefont {E.~R.}\ \bibnamefont {Hemesath}},
  \bibinfo {author} {\bibfnamefont {E.~J.}\ \bibnamefont {Schwalbach}},
  \bibinfo {author} {\bibfnamefont {J.~L.}\ \bibnamefont {Lensch-Falk}},
  \bibinfo {author} {\bibfnamefont {P.~W.}\ \bibnamefont {Voorhees}}, \ and\
  \bibinfo {author} {\bibfnamefont {L.~J.}\ \bibnamefont {Lauhon}},\
  }\href@noop {} {\bibfield  {journal} {\bibinfo  {journal} {Nature
  Nanotechnology}\ }\textbf {\bibinfo {volume} {4}},\ \bibinfo {pages} {315}
  (\bibinfo {year} {2009})}\BibitemShut {NoStop}%
\bibitem [{\citenamefont {Stamplecoskie}\ \emph {et~al.}(2008)\citenamefont
  {Stamplecoskie}, \citenamefont {Ju}, \citenamefont {Farvid},\ and\
  \citenamefont {Radovanovic}}]{Radovanovic}%
  \BibitemOpen
  \bibfield  {author} {\bibinfo {author} {\bibfnamefont {K.~G.}\ \bibnamefont
  {Stamplecoskie}}, \bibinfo {author} {\bibfnamefont {L.}~\bibnamefont {Ju}},
  \bibinfo {author} {\bibfnamefont {S.~S.}\ \bibnamefont {Farvid}}, \ and\
  \bibinfo {author} {\bibfnamefont {P.~V.}\ \bibnamefont {Radovanovic}},\
  }\href@noop {} {\bibfield  {journal} {\bibinfo  {journal} {Nano Lett.}\
  }\textbf {\bibinfo {volume} {8}},\ \bibinfo {pages} {2674} (\bibinfo {year}
  {2008})}\BibitemShut {NoStop}%
\end{thebibliography}%

\end{document}